\documentclass[pre,aps,twocolumn,10pt]{revtex4-2}
\usepackage{amsmath,amsfonts}
\usepackage{graphicx}
\usepackage{amssymb}
\usepackage{enumerate}
\usepackage{dcolumn}
\usepackage{textcomp}
\usepackage[colorlinks,citecolor=blue,linkcolor=blue]{hyperref}
\usepackage{natbib}
\usepackage{etoolbox}
\usepackage{mathrsfs}
\usepackage{bm}

\newcommand{\note}{\color{black}}
\begin{document}

\title{Sticky eigenstates in systems with sharply divided phase space}
\author{Hua Yan}
\email{yanhua@ustc.edu.cn}
% \author{Marko Robnik}
% \email{Robnik@guest.uni-mb.si}
\affiliation{Center for Applied Mathematics and Theoretical Physics,    University of Maribor, Mladinska 3, SI-2000 Maribor, Slovenia}
\date{\today} 
\begin{abstract}
We investigate mixed eigenstates in systems with sharply divided phase space, using different piecewise-linear maps whose regular-chaotic boundaries are formed by marginally unstable periodic orbits (MUPOs) or by quasi-periodic orbits. With the overlap index and the entropy localization length, we classify mixed eigenstates and show that the contribution from dynamical tunneling scales as $\sim \hbar\, \exp(-b/\hbar)$, with $b>0$ associated with the relative size of the regular region. The dominant fraction of states that remain sticky to the boundaries, referred to as sticky eigenstates, scales approximately as $\hbar^{1/2}$ in the MUPO case and oscillates around this algebraic behavior in the quasiperiodic case. This behavior generalizes established predictions for hierarchical states in Kolmogorov-Arnold-Moser (KAM) systems, which scale as $\hbar^{1 - 1/\gamma}$, with $\gamma$ set by the corresponding classical stickiness reflected in the algebraic decay of cumulative recurrence-time distributions $t^{-\gamma}$. For the piecewise-linear maps studied here, $\gamma = 2$. These results reveal a clear quantum signature of classical stickiness in non-KAM systems.
\end{abstract}  
\maketitle

\section{Introduction}
Rooted in the correspondence principle, long recognized as a central pillar of Bohr's philosophical interpretation of quantum mechanics, the principle of uniform semiclassical condensation (PUSC) of eigenstates asserts that  in the semiclassical limit almost all eigenstates  are either regular or chaotic \cite{percival1973regular,berry1977regular,robnik1998topics}. Their phase-space representations condense onto regions exhibiting regular or chaotic classical dynamics in systems with compact, mixed phase space. The PUSC further implies that chaotic eigenstates delocalize over the chaotic component in accordance with random matrix theory \cite{gutzwiller2013chaos,haake1991quantum,stockmann2007quantum}, whereas regular eigenstates remain confined to quantized invariant tori within the regular region, giving rise to Berry–Robnik spectral statistics \cite{berry1984semiclassical,prosen1994semiclassical,yan2025spacing}. Although this provides a qualitative picture of asymptotic behavior of eigenstates, understanding how quantum eigenstates approach this limit is considerably more challenging. Multiple mechanisms intervene, including dynamical tunneling \cite{bohigas1993manifestations,tomsovic1994chaos,keshavamurthy2011dynamical}, which refers to classically forbidden transitions between phase-space regions separated by dynamical barriers, and slow relaxation processes in classical transport that give rise to dynamical localization and delay the onset of quantum chaos \cite{izrailev1988quantum,chirikov1988quantum,batistic2013quantum}.

Extensive works have deepened the understanding of dynamical tunneling, using approaches such as random matrix models and the construction of fictitious integrable systems \cite{podolskiy2003semiclassical,backer2008regular,brodier2001resonance}. These studies reveal a tunneling rate scale as $\hbar\exp(-a/\hbar)$ with $a>0$ associated with the relative size of the regular region, for the direct regular-to-chaotic tunneling, while the nonlinear resonances within the regular region may significantly enhance this tunneling process. Concerning classical transport, particularly in the vicinity of the boundary between regular and chaotic regions, studies of mixed Kolmogorov-
Arnold-Moser (KAM) systems show that cantori-induced stickiness yields a power-law decay $Q(t)\sim t^{-\gamma}$ of the cumulative recurrence-time distributions (RTDs) \cite{altmann2013leaking}. {\note This indicates that chaotic trajectories can remain trapped near KAM surfaces for long times before escaping, thereby producing intermittent bursts of chaos motion characteristic of weak chaos.} The corresponding exponent displays a universal character: for bounded phase spaces it typically satisfies $1.5 \le \gamma \le 3$, whereas in unbounded phase spaces one finds $\gamma \simeq 1.5$ \cite{meiss1985markov,cristadoro2008universality,venegeroles2009universality}.

% Based on studies of the standard map, it has further been proposed that in KAM systems, this classical stickiness induces the formation of hierarchical states localized near cantori-mediated hierarchical structures at the regular–chaotic border, with a fraction that scales as  $\hbar^{1-1/\gamma}$ \cite{ketzmerick2000new}. 

{\note One quantum signature of universal power-law trapping is fractal conductance fluctuations in ballistic cavities with mixed phase space, such as two-dimensional billiards coupled to two infinite leads, whose fractal dimension is semiclassically derived as $D=2-\gamma/2$ \cite{ketzmerick1996fractal,alhassid2000statistical,huckestein2000quantum}, A second quantum signature, identified in studies of the standard map, is the formation of hierarchical states localized near cantori-mediated structures at the regular--chaotic border. In KAM systems, the fraction of such states is predicted to scale as $\hbar^{1-1/\gamma}$ \cite{ketzmerick2000new}. It remains unclear whether an analogous phenomenon occurs in non-KAM systems, such as piecewise-linear maps whose regular--chaotic boundaries are formed by marginally unstable periodic orbits (MUPOs) or quasiperiodic orbits. In these systems the phase space is sharply divided, similar to that of mushroom billiards \cite{bunimovich2001mushrooms,altmann2005stickiness,backer2008dynamical,barnett2007quantum,dettmann2011open,orel2026quantum}.} The boundaries between distinct dynamical regions are simple curves, rather than the hierarchical structures typical in KAM systems \cite{altmann2006stickiness,altmann2007intermittent,akaishi2009accumulation}. 

Recently, the concept of mixed eigenstates has provided a quantitative means to study eigenstates whose Husimi functions overlap both the regular and chaotic regions of phase space \cite{lozej2022phenomenology,yan2024chaos,yan2024further,wang2023power}. This measure, referred to as the overlap index, characterizes the degree to which the phase-space representation of an eigenstate spreads across the two components. Moreover, it has been found that in KAM systems the fraction of mixed eigenstates decays algebraically  with the (effective) Planck constant, as $\hbar^{\eta}$,  with $\eta>0$. It is natural to expect a connection between mixed eigenstates and hierarchical states, since the finite width of the coherent states underlying the Husimi representation tends to smooth or blur hierarchical eigenstates across the regular–chaotic border. 

In this paper, we take advantage of the overlap index and combine it with another measure, the entropy localization length, to address both aspects of the PUSC picture, namely the degree of concentration and the degree of localization. This allows us to classify mixed eigenstates in non-KAM systems and thereby identify sticky eigenstates. In particular, we show that their fraction exhibits a pronounced power-law decay, with exponents close to those extracted from RTDs. This combined approach allows us to detect a clear quantum signature of classical stickiness in systems with sharply divided phase space, generalizing predictions for hierarchical states in KAM systems such as the standard map. {\note  The sharply divided phase space of the system considered here provides a simple and transparent setting for examining how phase space transport structures influence quantum eigenstates. From this perspective, the results may offer useful insight into broader questions concerning the mechanisms behind nonthermal eigenstates, scar-like localization, and deviations from full ergodicity in many-body systems \cite{serbyn2021quantum,moudgalya2022quantum}.} 

The paper is organized as follows. In Sec.~\ref{sec:sec2}, we analyze the classical dynamical features of several piecewise-linear maps with sharply divided phase space, as well as their canonical quantization on the torus. In Sec.~\ref{sec3}, we introduce the overlap index and the entropy localization length, and use them to identify and characterize the sticky eigenstates. In Sec.~\ref{sec:rmt}, using random-matrix modeling, we reveal the decay behavior of mixed eigenstates originating from dynamical tunneling. Finally, we present our conclusions in Sec.~\ref{sec:sum}. Details on the RTDs are provided in the Appendix \ref{app:rts}.

\section{Piecewise-linear maps and their quantization}
\label{sec:sec2}
\subsection{Piecewise-linear symplectic maps}
\label{sec:sec2a}
We consider area-preserving, discrete-time classical maps $\Omega$ on a phase space with the topology of a two-dimensional torus $\mathbb{T}^2$, given by the form
\begin{equation}
    \begin{aligned}
        p_{n+1}&=p_n+kf(x_n) \quad (\text{mod}\ 1),\\
        x_{n+1}&=x_n+p_{n+1}\quad (\text{mod}\ 1),
    \end{aligned}
\end{equation}
where $k$ acts as a control parameter that determines the transition from regular to chaotic (ergodic) dynamics of the map. When $f(x_n)=\sin(2\pi x_n)$, the system is the standard map, a prototype model widely used to study the hierarchical structure of KAM islands and cantori. Following Ref. \cite{lee1989makes,malovrh2002spectral,ishii2012weyl}, we consider two different maps: (i) the continuous sawtooth map (distinct from the hyperbolic discontinuous sawtooth map \cite{devaney2018introduction,dana1989resonances,vaienti1992ergodic}), where
\begin{align}
    f(x_n)=1-|2x_n-1|,
\end{align}
and (ii) 
\begin{align}
    f(x_n) = 
\begin{cases}
    -x_n & \text{if } 0 \leq x_n < \frac{1}{4}, \\
    -\frac{1}{2} + x_n & \text{if } \frac{1}{4} \leq x_n < \frac{3}{4}, \\
    1 - x_n & \text{if } \frac{3}{4} \leq x_n \leq 1.
\end{cases}
\end{align}
\begin{figure}[h]
    \centering
    \includegraphics[width=1\linewidth]{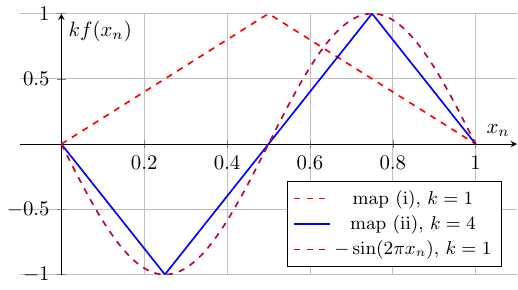}
    \caption{Illustration of two piecewise-linear functions alongside the standard map for comparison.}
    \label{fig:scheme}
\end{figure}

These two maps can be viewed as linearized versions of the standard map, with the piecewise-linear function $f(x_n)$ approximating the sine function, as shown in Fig.~\ref{fig:scheme}. {\note In both cases, \(f(x)\) can be represented through the Taylor expansion as the limit of a truncated smooth series $f(x) = \lim_{M\to\infty}f_M(x)$. For map (i), this gives
\begin{align}
\label{eq:map1-smooth}
  f_M(x) = \frac{2}{\pi}\sum_{l=0}^M \frac{(2l-1)!!}{(2l)!!(2l+1)}\sin^{2l+1}(\pi x),
\end{align}
whereas for map (ii), one obtains
\begin{align}
\label{eq:map2-smooth}
  f_M(x) = -\frac{1}{2\pi}\sum_{l=0}^M \frac{(2l-1)!!}{(2l)!!(2l+1)}\sin^{2l+1}(2\pi x).
\end{align}

\begin{figure*}
    \includegraphics[width=0.9\linewidth]{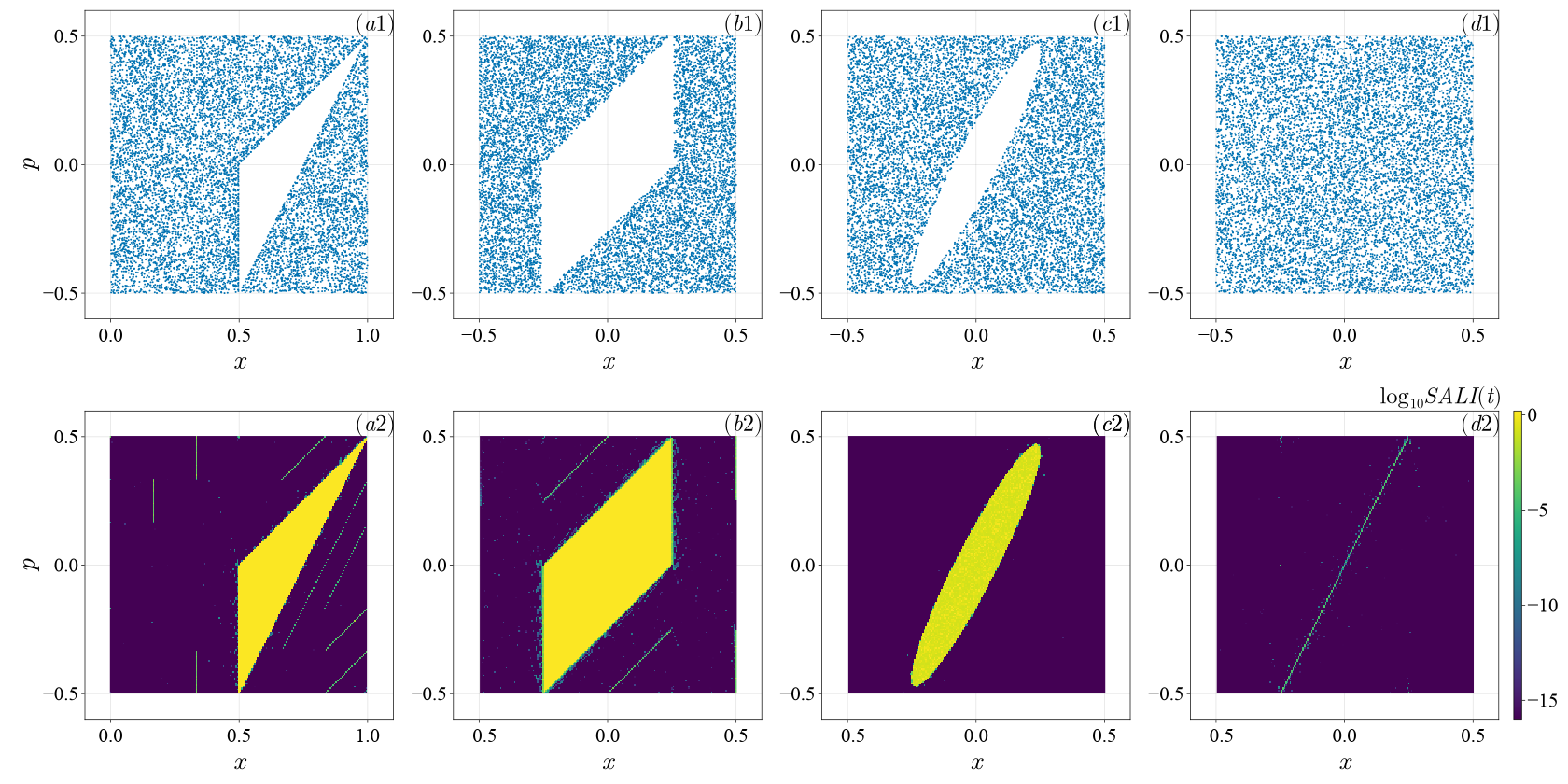}
        \caption{Comparison between the Poincaré sections in panels (a1)–(d1) and the SALI plots in panels (a2)–(d2), for maps (i) with $k = 1.5$ and (ii) with $k = 2,3.5,4$, from left to right. Panels (a1)-(d1) show a single chaotic orbit evolved for $10^4$ iterations. The SALI values, shown on a logarithmic scale in the lower panels, are computed up to 60 iterations. We plot $0\le x\le  1, - 0.5\le p \le 0.5$ in (a) and $ - 0.5 \le x,p \le 0.5$ in (b)-(d) for visualization convenience.}
    \label{fig:cp1}
\end{figure*}

For small values of \(M\), the smoothing of \(f_M(x)\) induces tiny island chains around the main regular island, as expected from the KAM theorem. As \(M\) increases, these island chains generally become less visible as the border between regular and chaotic regions becomes sharper, in the limit \(M\to\infty\), they disappear, as shown in the Appendix.~\ref{app:island}. This occurs because the limiting potential \(f(x)\) violates the smoothness assumptions of the KAM theorem in a minimal way: It is a \(C^1\), but not a \(C^2\) function of the coordinate.} As a result, the phase space can be sharply divided, meaning that regular and chaotic regions of positive measure are separated by non-hierarchical borders formed by simple curves, in contrast to the hierarchical structures commonly found in generic Hamiltonian systems. These borders can arise from MUPOs or quasi-periodic orbits, along which sticky motion occurs. In such cases, especially those involving MUPOs, the cumulative RTD generally follows a power-law $\sim t^{-\gamma}$ where $\gamma=2$ \cite{altmann2006stickiness,altmann2007intermittent,akaishi2009accumulation} (see numerical results and discussions in Appendix~\ref{app:rts}).

For $k>2$ map (i) is uniformly hyperbolic, since then all periodic orbits are hyperbolic. It was also proved that \cite{wojtkowski1981model} map (ii) is fully chaotic and there are no stable islands in the phase space for $k\ge 4$, while for $k<4$ there is an isolated island of regular motion with elliptical shape if 
\begin{align}
    \frac{1}{\pi}\arccos(\frac{k}{2}-1)\in  \mathbb{R} \setminus \mathbb{Q}
    \label{eq:cond}
\end{align}
is an irrational number and a polygonal shape otherwise. More specifically $k = 2(1+\cos \frac{\pi}{n})$ $(n=2,3,\cdots)$, a single stable island appears in the form of a $2n$-sided polygon, composed of periodic orbits of period $n$.

For both maps, Fig.~\ref{fig:cp1} compares the Poincaré sections with the smaller alignment index (SALI) plots~\cite{skokos2016chaos}. As an effective chaos indicator, SALI tracks the evolution of two random deviation vectors $\mathbf{w}$ along an orbit, governed by the tangent map $\mathbf{w}_{n+1} = \left( \partial \Omega / \partial \mathbf{x}_n \right) \cdot \mathbf{w}_n$, where $\mathbf{x}_n = (x_n, p_n)$. In Hamiltonian systems with toroidal phase space, a chaotic orbit typically exhibits a single positive Lyapunov exponent \(\lambda_1\), and it can be proven that SALI$(t)\propto e^{-\lambda_1t}$. In the case of regular motion, however, SALI$(t) \sim const.>0$ as $t\to \infty$.  SALI plots reveal fine structures, similar to finite-time Lyapunov exponents (FTLEs), highlighting the intermittent dynamics along the boundaries between chaotic and regular regions formed by MUPOs with null stability exponent, as well as MUPOs within the chaotic region where stickiness appears as intermediate colors.

In the continuous sawtooth map for $k=1.5$ [see Fig.~\ref{fig:cp1}(a2)], families of period-three MUPOs appear within the chaotic region, namely the sets $\{x=\frac{1}{6}, \frac{1}{6} \le p\le \frac{1}{3}\}$,  $\{x=\frac{1}{3}, \frac{1}{3}\le p\le \frac{2}{3}\}$ and their images,
as well as MUPOs along the island border, both indicated by intermediate colors. In map (ii), at \( k=2 \) [Fig.~\ref{fig:cp1}(b2)], families of MUPOs lie along the boundary of the polygonal regular island and also within the chaotic sea, from two sets $\{x=p+0.5, 0\le x\le 0.25\}$, $\{x=0.5, 0.25\le y\le 0.5\}$, together with their image. At \( k=4 \) [Fig.~\ref{fig:cp1}(d2)] there are no stable
islands in phase space, and only one family of MUPOs of measure zero remains in the fully chaotic phase space. In particular, on \( k=3.5 \) on map (ii) (Fig.~\ref{fig:cp1}(c2)), the isolated island is elliptical, as predicted by Eq.~\eqref{eq:cond}, but its border is formed by a quasi-periodic trajectory rather than MUPOs, and stickiness persists, consistent with Ref.~\cite{altmann2007intermittent}, and there are no MUPOs in the chaotic region.

\begin{figure*}
    \centering
    \includegraphics[width=0.9\linewidth]{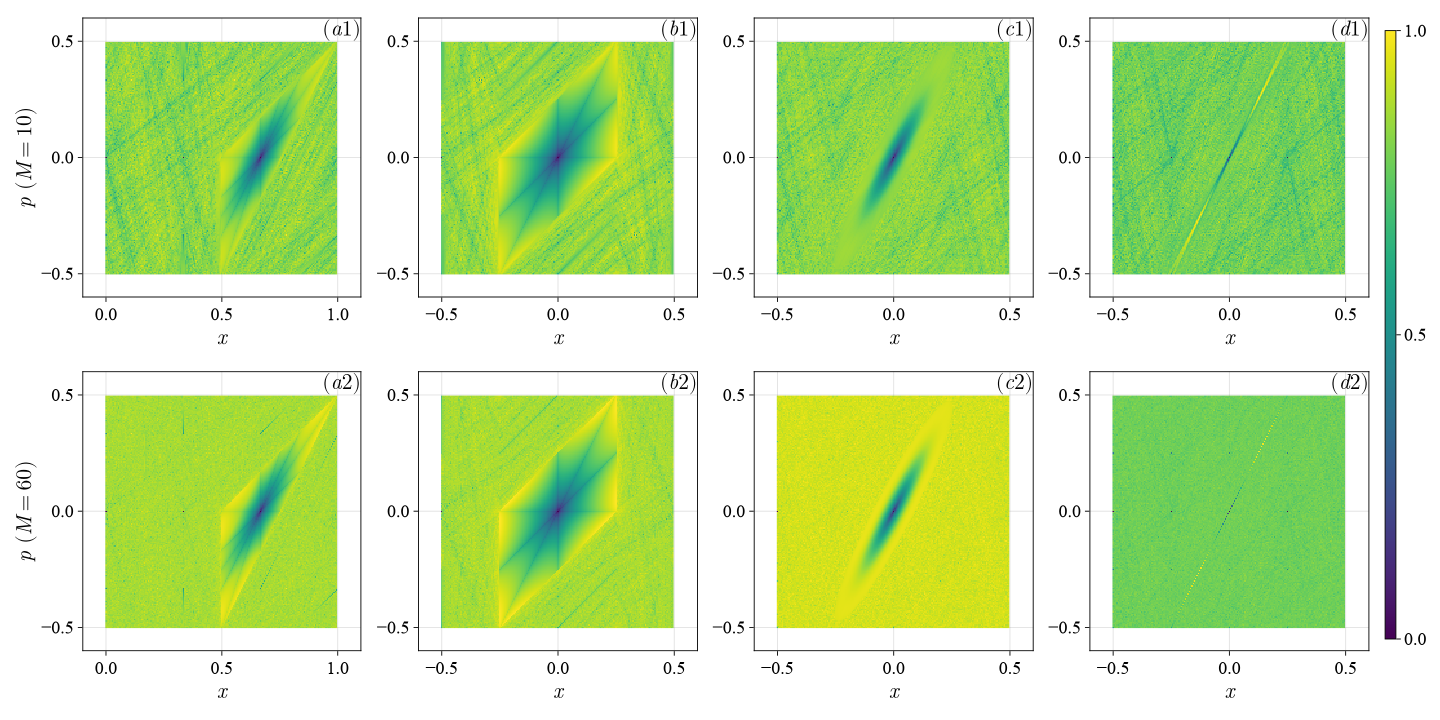}
    \caption{Lagrangian descriptors for the two maps with $a=0.5$, shown from left to right and normalized to their respective maximum LD values, are presented as in Fig.~\ref{fig:cp1}, where the top and bottom panels correspond to integration times $M=10$ and $M=60$, respectively.}
    \label{fig:ld}
\end{figure*}

\subsection{Lagrangian descriptors}
Lagrangian descriptors (LDs) were originally introduced as a tool to identify hyperbolic trajectories (saddle points). More recently, they have been used as a classical measure to characterize the phase-space structure of dynamical systems \cite{mancho2013lagrangian,lopesino2015lagrangian}, particularly the stable and unstable manifolds of hyperbolic invariant sets associated with the unstable periodic orbits (UPOs), such as homoclinic tangles, since the derivative of the LD becomes unbounded across these manifolds. It should be noted that LDs can also be used to determine whether orbits in a system’s phase space are chaotic or regular, based on the difference or ratio of the LDs for those orbits and their nearby trajectories \cite{hillebrand2022quantifying}. For area preserving maps, the LD corresponding to an orbit $\{x_i,p_i\}_{i=-M}^{i=M-1}$  of length $2M+1$ generated by the maps is defined as
\begin{align}
    LD_a=LD_a^+ +LD_a^-,
\end{align}
or, in a slightly modified form,
\begin{align}
         LD_a=LD_a^+ \times LD_a^-,
 \end{align}
 with $a\le 1$ and normalizing with respect to the maximal value of LDs in the phase space, where the forward and backward contributions are given by
\begin{align}
    LD_a^+= \sum_{i=0}^{M-1}|x_{i+1}-x_i|^a+|p_{i+1}-p_i|^a,
\end{align}
and 
\begin{align}
    LD_a^-= \sum_{i=-M}^{-1}|x_{i+1}-x_i|^a+|p_{i+1}-p_i|^a.
\end{align}

The minimal values of the LD naturally occur at fixed points, as shown in Fig.~\ref{fig:ld}: for map (i) at $(0,2/3)$ and for map (ii) at $(0,0)$, the latter being independent of $k$. Both points lie within the regular island, as revealed by the Poincar\'e and SALI plots in Fig.~\ref{fig:cp1}. The ridges or discontinuities of LD field indicate the stable and unstable manifold, or the transport barriers. In particular, the short-time LD field $(M=10)$ shown in Fig.~\ref{fig:ld}(a1–d1) reveals structures within the chaotic region associated with the stable and unstable manifolds of the UPOs. In contrast, the longer time LD field $(M=60)$ shown in Fig.~\ref{fig:ld}(a2–d2) highlights MUPOs in the chaotic sea, as well as those serving as transport barriers at the border between the chaotic sea and the regular island, in agreement with the regions shown in intermediate colors in the SALI plots.

\subsection{Canonical quantization and quantum maps}
The phase space of the classical maps is a two-torus, so quantization imposes the same periodicity on quantum states. As a result, quantum states are periodic up to phase factors under translations by the fundamental periods of the torus in the position and momentum representations \cite{leboeuf1990phase,saraceno1994towards,keating1999quantum,hannay1980quantization}, that is, the wavefunction $\psi(x)$ and $\hat{\psi}(p)$ satisfy
\begin{align}
    \psi(x+1)=e^{2\pi i\theta_2}\psi(x), \quad \hat{\psi}(p+1)=e^{-2\pi i\theta_1}\hat{\psi}(p),
\end{align}
resulting in the quantization of $x_j= j/N$ and $p_k=k/N$
for $j,k \in \mathbb{N}$. The invariance imposed on the Hilbert space in both translations implies that the finite dimension of the Hilbert space should satisfy $N=(2\pi\hbar)^{-1}$. The semiclassical limit $\hbar\to 0$ thus corresponds to $N\to\infty$. The phases $\bm{\theta} =(\theta_1,\theta_2)\in[0,1]^2$ are at first arbitrary, label a family of Hilbert space $\mathcal{H}_N(\bm{\theta})$, and one obtains the periodic boundary conditions for $\theta_1=\theta_2=0$, which we will use in the following. A state $|\psi\rangle$ of $\mathcal{H}_N$ admits a natural position representation as a periodical sum
\begin{align}
\label{eq:torus-state}
    \psi(x)=\sum_{\nu=-\infty}^\infty\sum_{j=0}^{N-1}\psi_j\delta(x-j/N-\nu),
\end{align}
where $\psi_j=\langle x_j|\psi\rangle$ and $|x_j\rangle$ being the state localized at $x=x_j(\text{mod}\ 1)$, form the complete basis of discrete position states. 

One can build an analytical coherent state representation of $\mathcal{H}_N$ from the standard Weyl group coherent states on the plane $\mathbb{R}^2$, that $|\alpha\rangle=\exp(\alpha a^\dagger-\alpha^* a)|0\rangle$ and 
\begin{align}
\label{eq:kernel}
    \langle x|\alpha\rangle =\frac{e^{-|\alpha|^2/2\hbar}}{(\pi\hbar)^{1/4}}\exp\{-[\frac{1}{2}(\alpha^2+x^2)-\sqrt{2}\alpha x]/\hbar\},
\end{align}
where $\alpha=(x+ip)/\sqrt{2}$.  By introducing Eq.~\eqref{eq:kernel} into Eq.~\eqref{eq:torus-state} together with the consistency condition $N=(2\pi\hbar)^{-1}$, the state $|\psi\rangle$ of $\mathcal{H}_N$ can be decomposed by coherent states on the torus as
\begin{align}
    \langle\alpha|\psi\rangle&=\int_{-\infty}^\infty dx \langle \alpha|x\rangle \psi(x)=\sum_{j=0}^{N-1}\langle\alpha|x_j\rangle\psi_j,
\end{align}
where the coherent state in $\mathbb{T}^2$ in discrete position satisfies
\begin{align}
\label{eq:cs-torus}
    \langle \alpha|x_j\rangle
    %=\sum_{\nu=-\infty}^\infty\langle \alpha|x_j+\nu\rangle
    =(2N)^{\frac{1}{4}}e^{-2\pi N[(\bar{\alpha}^2+|\alpha|^2+x_j^2)/2-\sqrt{2}\bar{\alpha}x_j]}\nonumber \\
    \theta_3\left(i N[x_j-\sqrt{2}\bar{\alpha}]|iN\right),
\end{align}
and $\theta_3$ is the Jacobi-$\theta$ function 
\begin{align}
    \theta_3(z|\tau)=\sum_{\nu=-\infty}^\infty e^{i \pi  \nu^2\tau+2\pi i \nu z}, \text{\quad with}\ \text{Im}(\tau)>0.
\end{align}

The unitary operator $U_N$ in $\mathcal{H}_N$, which corresponds to the classical map, is determined by its generating function of that map \cite{hannay1980quantization,keating1991cat,backer2003numerical}
\begin{align}
    (U_N)_{j',j}&\equiv\langle x_{j'}|U_N|x_j\rangle\nonumber \\
    =&\frac{1}{\sqrt{N}}\left|\frac{\partial^2 S(x',x)}{\partial x'\partial x}\right|^{1/2}_{x_{j'},x_j}\!\!\!\!\exp(2\pi i NS(x_{j'},x_j)),
\end{align}
the classical map $\Omega$ is here denoted as $(x,p)\to(x',p')$ and the generating function gives $p=-\partial S(x',x)/\partial x$ and $p'=\partial S(x',x)/\partial x'$. It is easy to check that
\begin{align}
    S(x',x)= \frac{1}{2}(x-x')^2+kh(x),
\end{align}
where $f(x)=\partial h(x)/\partial x$. For map $(i)$
\begin{align}
    h(x) = 
\begin{cases}
   x^2+1/4, &  x<1/2, \\
    -x^2+2x-1/4, & x>1/2,
\end{cases}
\end{align}
and map $(ii)$
\begin{align}
    h(x) = 
\begin{cases}
    -\frac{1}{2}x^2, & 0 \leq x < \frac{1}{4}, \\
    \frac{1}{2}x^2-\frac{1}{2}x+1/16, &  \frac{1}{4} \leq x < \frac{3}{4}, \\
    x-\frac{1}{2}x^2-1/2,&  \frac{3}{4} \leq x \leq 1,
\end{cases}
\end{align}
the unitary operator $U_N$  reads 
\begin{align}
    (U_N)_{j',j}=\frac{1}{\sqrt{N}}\exp(2\pi iN[\frac{1}{2}(x_j'-x_j)^2+kh(x_j)]),
\end{align}
note that here the constant added to $h(x)$ is arbitrary and does not affect the spectra or eigenfunction statistics.

{\note The canonical quantization of piecewise-linear maps on the torus yields finite-dimensional Floquet models, in which the discrete Fourier transform connects the position and momentum bases and enables a construction from diagonal phase factors and Fourier-transform blocks. Examples such as the quantum baker map and the quantum sawtooth map, have been realized on quantum-information platforms including nuclear magnetic resonance and superconducting-qubit processors \cite{brun1999realizing,weinstein2002experimental,patoary2024chaotic,benenti2026quantum}}.

\section{Husimi function}
\label{sec3}
The Husimi function as the phase space representation of eigenstates allows for a direct comparison with the classical phase space structure. The quasienergy spectrum $\{\epsilon_n\}$ from diagonalization $U_N|\epsilon_n\rangle=\exp(-i\epsilon_n)|\epsilon_n\rangle$, while the Husimi representation is obtained by projecting the eigenvector onto a coherent state $|\alpha\rangle$ centered at $(x,p)\in\mathbb{T}^2$, as defined in Eq.~\eqref{eq:cs-torus}
\begin{align}
    H_n(x,p)=\frac{1}{N}|\langle\alpha|\epsilon_n\rangle|^2=\frac{1}{N}\bigg|\sum_{j=0}^{N-1}c_{nj}\langle \alpha|x_j\rangle\bigg|^2,     
\end{align}
where $c_{nj}\equiv\langle x_j|\epsilon_n\rangle$ are the coefficients of $|\epsilon_n\rangle$ in the position basis. The prefactor $1/N$ accounts for the fact that the coherent state $|\alpha\rangle$ is not normalized, with $\langle\alpha|\alpha\rangle=N$. 

\begin{figure}[h]
    \centering
    \includegraphics[width=1.0\linewidth]{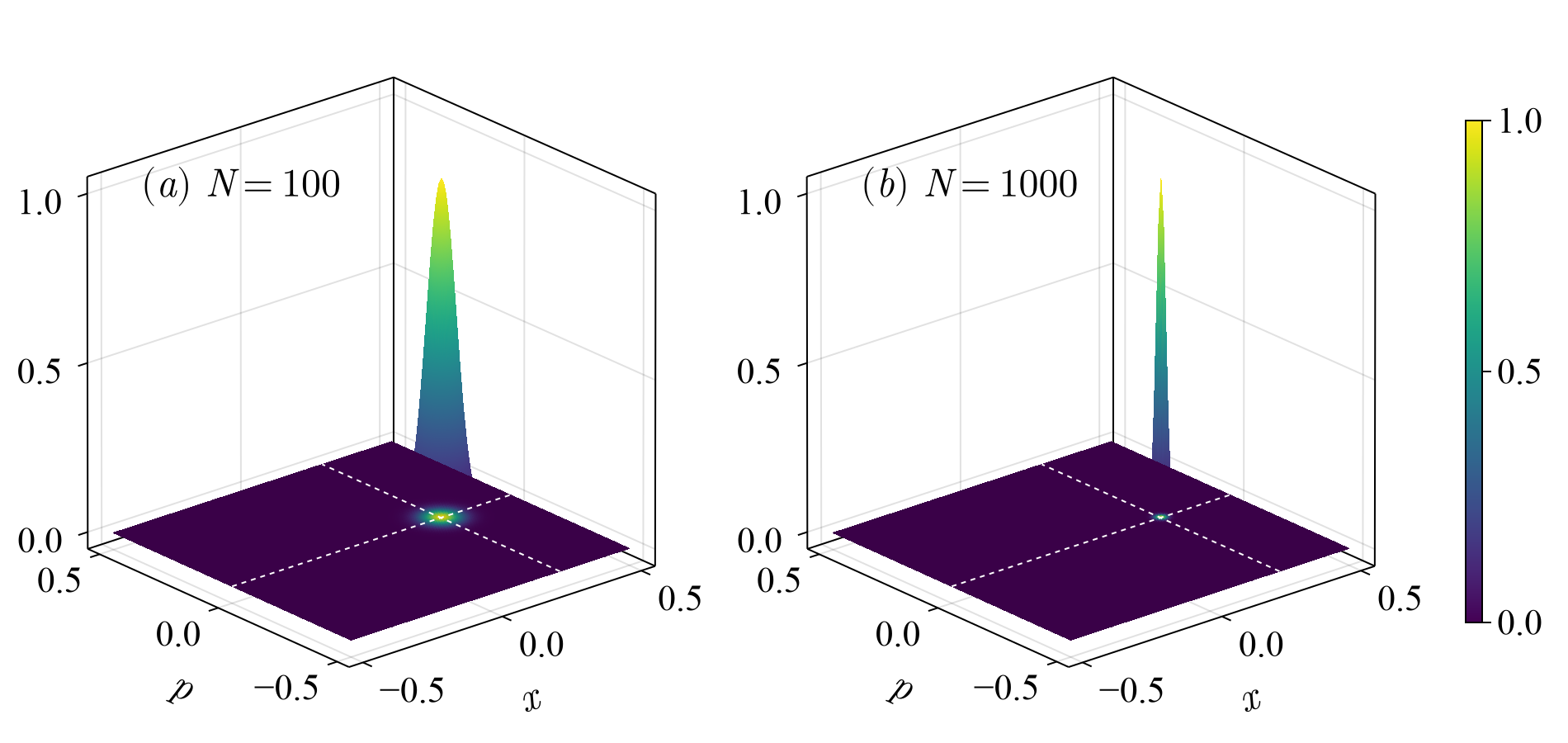}
    \caption{Three-dimensional representation of the normalized coherent state $|\alpha\rangle$ centered at $(0.25,0)$, showing the surface together with its projection onto the $(x,p)$-plane, for system sizes $N=100$ in (a) and $N=1000$ in (b).}
    \label{fig:coherent_bump}
\end{figure}

\begin{figure*}
    \centering
    \includegraphics[width=0.9\linewidth]{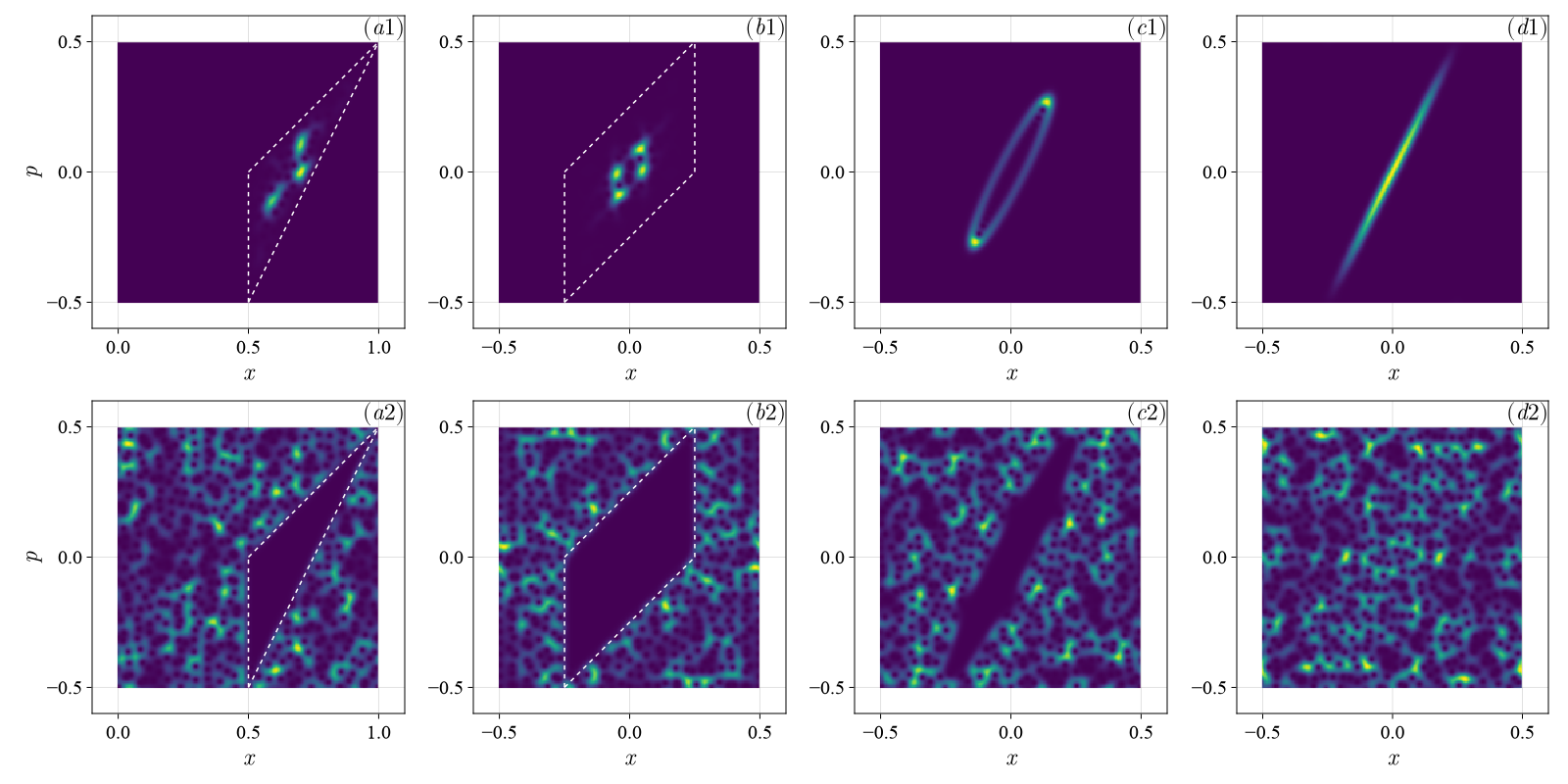}
    \caption{Examples of Husimi functions $H_n(x,p)$ for two maps, presented in the same manner as Figs.~\ref{fig:cp1}-\ref{fig:ld}, with $N=500$. The upper panels (a1–c1) correspond to eigenstates with  overlap index $\omega \simeq -1$, while the lower panels (a2-c2) show states with $\omega \simeq 1$. All data are shown in a linear scale. Bright colors indicate regions of large $H_n(x,p)$, and dashed lines in panels (a) and (b) mark the boundaries between regular and chaotic regions.}
    \label{fig:husimi-maps}
\end{figure*}

For $N\gg1$, we show in Appendix~\ref{app:asymp} that the coherent state on the torus admits the asymptotic form 
\begin{align}
\label{eq:asy-torus}
   \langle x_j|\alpha\rangle\simeq\exp(-\pi Nr_j^2)\exp\left(-i\pi Np(x+2r_j)\right),
\end{align}
where $r_j=x_j-x - \lfloor x_j-x+\frac{1}{2}\rfloor$, which is equivalent to stating that $|r_j|=\text{dist}(x-x_j, \mathbb{Z})$, i.e., the distance from $x-x_j$ to the nearest integer. The coherent state, appearing as a Gaussian bump in phase space, is the minimum uncertainty state that saturates the Heisenberg bound $\Delta x\Delta p \sim \hbar$ when expressed in dimensionless variables satisfying $\Delta x = \Delta p$. For a coherent state on the torus, this relation becomes $\Delta x \Delta p\sim  1/(2\pi N)$, as seen from Eq.~\eqref{eq:asy-torus}, the Gaussian form has the variance $\Delta x= 1/\sqrt{2\pi N}$. Consequently, the bump becomes narrower as the system size $N$ increases, as shown in Fig.~\ref{fig:coherent_bump}.

The asymptotic form of the coherent state on torus allowing the Husimi function to be simplified as
\begin{align}
\label{eq:husimi-bound}
    H_n(x,p) \simeq (\frac{2}{N})^\frac{1}{2}\bigg|\sum_{j=0}^{N-1}c_{nj}e^{-\pi Nr_j^2}e^{-i\pi Np(x+2r_j)}\bigg|^2.
\end{align}
In systems with mixed phase space, the Husimi functions of eigenstates are located in either regular or chaotic regions in the semiclassical limit, provided that dynamical tunneling between classically disconnected regions is negligible, in accordance with the PUSC. To better quantify the concentration in each region, we define the overlap index of eigenfunctions \cite{batistic2013dynamical,robnik2023recent} as
\begin{align}
\label{eq:olap-map}
    \omega_n&=\frac{1}{2\pi\hbar}\int H_n(x,p)\chi(x,p) dxdp \nonumber \\&= N\int H_n(x,p)\chi(x,p) dxdp,
\end{align}
where the Husimi functions are normalized such that $N\int H_n(x,p)dxdp=1$. Here $\chi(x,p)$ is the characteristic function takes the value of +1 in the chaotic region and -1 in the regular. Thus, $\omega_n$
provides a measure of how strongly the Husimi distribution of the eigenstate is concentrated in either region. Fig.~\ref{fig:husimi-maps}(a-c) shows the Husimi functions of two maps with overlap index close to two limiting values $\omega =\pm 1$, where for eigenstates with $\omega \simeq 1$ we select those that are extended in the available phase space (not the localized chaotic eigenstates). Fig.~\ref{fig:husimi-maps}(d1-d2) corresponds to the case containing only one family of MUPOs (of measure zero) in the phase space: the upper panel shows the eigenstate that is strongly localized around the MUPOs, while the lower panel displays the chaotic eigenstate that is delocalized across all the available phase space.

Besides the eigenstates with overlap index $\omega \simeq \pm 1$, there exist eigenstates with intermediate values of $\omega$, which we refer to as \emph{mixed eigenstates}. To quantify the fraction, we define
\begin{equation}
f_M = \frac{1}{N} \sum_{n=0}^{N-1} \mathbf{1}_{\{|\omega_n| \le \omega_*\}},
\end{equation}
where the indicator function $\mathbf{1}_{\{|\omega_n| \le \omega^*\}} = 1$ when $|\omega_n| \le \omega^*$ and $0$ otherwise. The threshold $\omega^*<1$ specifies the window defining mixed eigenstates.

\subsection{Mixed eigenstates}
In mixed KAM systems, the primary regular island is surrounded by chains of resonant sub-islands, producing a nested hierarchical structure at the interface between the regular region and the surrounding chaotic sea. When invariant tori are destroyed, they break into cantori that serve as partial transport barriers within the chaotic region. In such systems, previous studies \cite{lozej2022phenomenology,yan2024chaos,yan2024further,wang2023power} have shown that the fraction of mixed eigenstates scales as $f_M \sim \hbar^{\eta}$ with $\eta>0$. It is therefore instructive to contrast this behavior with systems whose mixed phase space is of non-KAM type. Examples include mushroom billiards and piecewise-linear maps with sharply divided phase space, where the boundaries between regular and chaotic regions are simple smooth curves or lines, and neither hierarchical structures nor cantori arise. Instead, MUPOs or quasi-periodic trajectories generate the sticky boundaries, as shown in Appendix~\ref{app:rts}.

\begin{figure}
    \centering
    \includegraphics[width=1\linewidth]{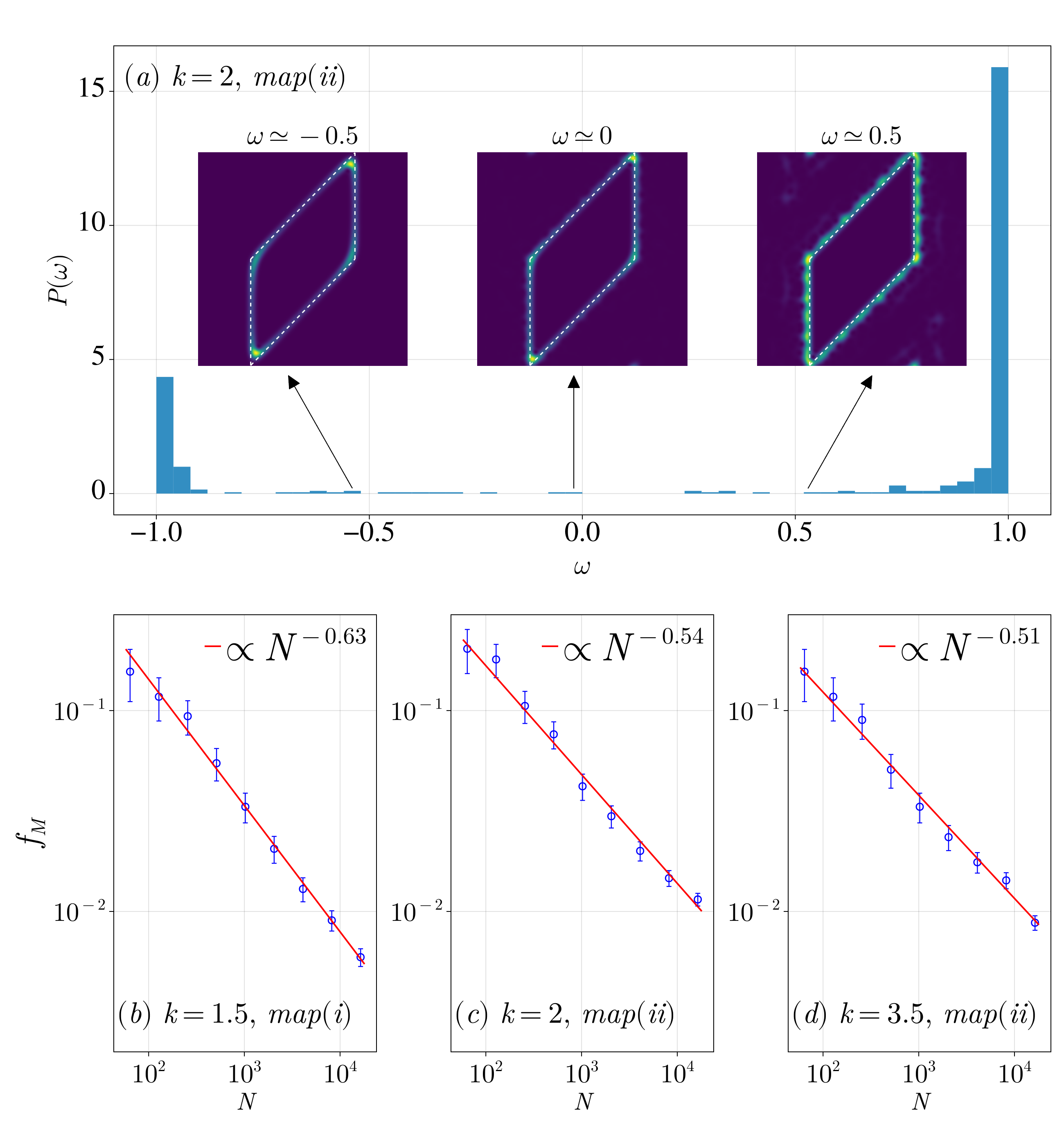}
    \caption{(a) Histogram of the overlap index $\omega$ for map (ii) at $k=2$, with examples of Husimi function (shown in linear scale) for three typical values $\omega=-0.5, 0, 0.5$, where the system size $N=500$. (b-d) Fraction of mixed eigenstates $f_M$ with $|\omega|\le 0.8$ as a function of $N =(2\pi\hbar)^{-1}$, for two maps at different values of $k$. Fits of the fractions are shown as solid lines, error bars reflect the uncertainties based on the number of states in each decade, assuming independent counts.}
    \label{fig:mixed}
\end{figure}

In these systems, the histogram of the overlap index in Fig.~\ref{fig:mixed}(a) clearly shows that mixed eigenstates persist. The Husimi functions for eigenstates with $\omega=-0.5$ and $0$ demonstrate states with substantial weight in the regular region, whereas the case $\omega=0.5$ illustrates an eigenstate localized near the MUPOs, carrying most of its weight in the chaotic sea. More importantly, Figs.~\ref{fig:mixed}(b–d) show that the fraction of mixed eigenstates in two quantum maps still follows a power-law decay, $f_M\sim \hbar^{\eta}$, though with different exponents $\eta$. Throughout this analysis we adopt a threshold $\omega^*=0.8$; other reasonable choices yield similar power-law behavior with only minor variations in the extracted exponents. Figures 6(b) and 6(c) correspond to maps whose sticky boundaries are formed by MUPOs, while Fig. 6(d) shows the case of map (ii), where an elliptic quasi-periodic orbit forms the boundary.

The example states with $\omega \simeq -0.5$ and 0 further reveal a subtlety in defining mixed eigenstates: the finite spatial extent of the coherent state (a Gaussian bump, shown in Fig.~\ref{fig:coherent_bump}) used to define the Husimi function can cause regular states localized near or on the boundary to be counted as mixed. This subtlety can be addressed more effectively by studying the localization properties of the mixed eigenstates.

\subsection{Localization of mixed eigenstates}
To resolve the subtlety noted above in defining mixed eigenstates, we supplement the overlap index with a characterization of the phase-space localization. For each eigenstate we compute the Wehrl entropy \cite{wehrl1979relation,lieb1978proof,gnutzmann2001renyi}
\begin{align}
    S_n= -N\int dxdp \ H_n(x,p)\ln H_n(x,p),
\end{align}
with $0 \le H_n(x,p)\le 1$ from Eq.~\eqref{eq:husimi-bound}, which implies bounds on $S_n$. For comparison, the mean Wehrl entropy of a random pure state is
\begin{align} 
    \langle S\rangle_N=\Psi(N+1)-\Psi(2)\sim \ln N+\gamma_e -1,
\end{align}
which equals the Shannon entropy of its expansion in a relatively random basis, where the expansion components are complex random numbers. Here, $\Psi(x)$ denotes the digamma function and $\gamma_e$ the Euler constant. 

\begin{figure}
    \centering
    \includegraphics[width=1.0\linewidth]{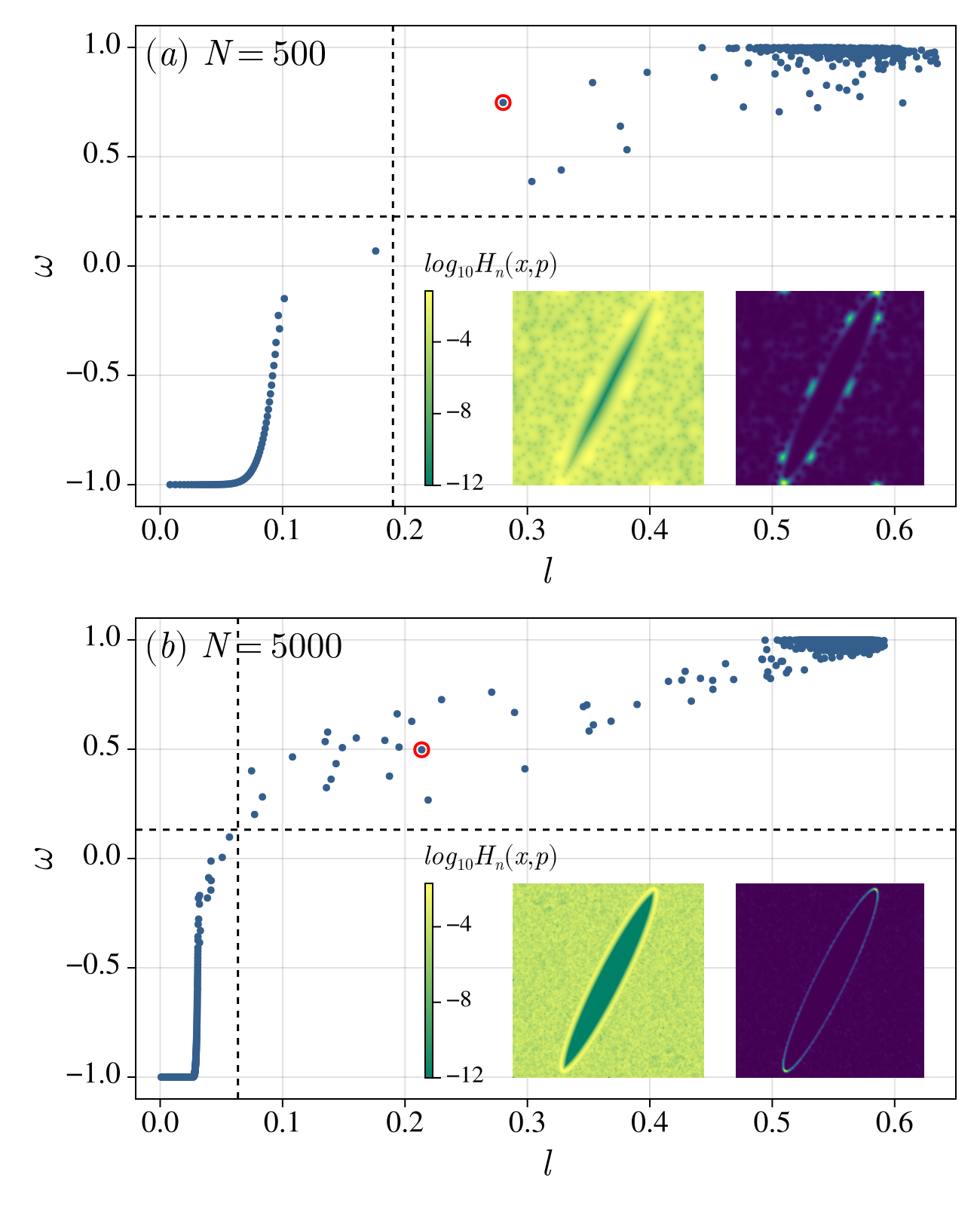}
    \caption{Joint distribution of entropy localization length $l$ and the overlap index $\omega$, for map (ii) at $k=3.5$ where no MUPOs exist in the chaotic region, with $N=500$ (a) and $N=5000$ (b). The inset shows the Husimi functions, plotted on both logarithmic and linear scales, corresponding to the eigenstates indicated by the red circles. For the logarithmic scale, a cutoff of $H_n(x,p)\le 10^{-12}$ is applied.  The vertical and horizontal dashed lines roughly partition the states into the lower-left and upper-right groups, where the number of states in the lower-left group is $NA_R$.}
    \label{fig:map-joint-dis}
\end{figure}

Based on the Wehrl entropy, we define the entropy localization length as 
\begin{align}
  l_n={\exp(S_n)}/{N}.  
\end{align}
 Different mechanisms of eigenstate localization in quantum chaotic systems, which in the classical limit are of weak or strong ergodicity, lead to deviations from RMT predictions. These mechanisms include quantum scarring caused by short unstable periodic orbits and strong scarring associated with MUPOs (or bouncing-ball modes in quantum billiards), and the interplay between two typical time scales: the classical transport time that characterizes the classical ergodicity, and the Heisenberg time that it takes to resolve individual eigenstates. Therefore, the maximum entropy localization length of the eigenstates, approximated by the mean value of a random pure state, is $l_{\max} \simeq e^{\gamma_e-1} \simeq 0.655$.

 In mixed-type systems, the PUSC suggests that chaotic eigenstates are delocalized over the chaotic region following the RMT, whereas regular eigenstates are localized on the quantized tori within the regular region. Yet this picture is incomplete without considering tunneling between chaotic and regular states and the role of eigenstates sticky around the boundary. {\note These mechanisms describe and quantify the approach to the PUSC scenario which always appears in the strict semiclassical limit.} A joint distribution of the entropy localization length and the overlap index is well suited for examining both aspects of the PUSC picture, namely the degree of concentration and the degree of localization. To address this issue more accurately, we chose map (ii) with $k=3.5$ in which the regular region of elliptical shape has relative size $A_R\simeq 0.13$ and no MUPOs appear in the chaotic region, thus eliminating strong scarring by MUPOs on chaotic eigenstates.

\begin{figure}
    \centering
    \includegraphics[width=1\linewidth]{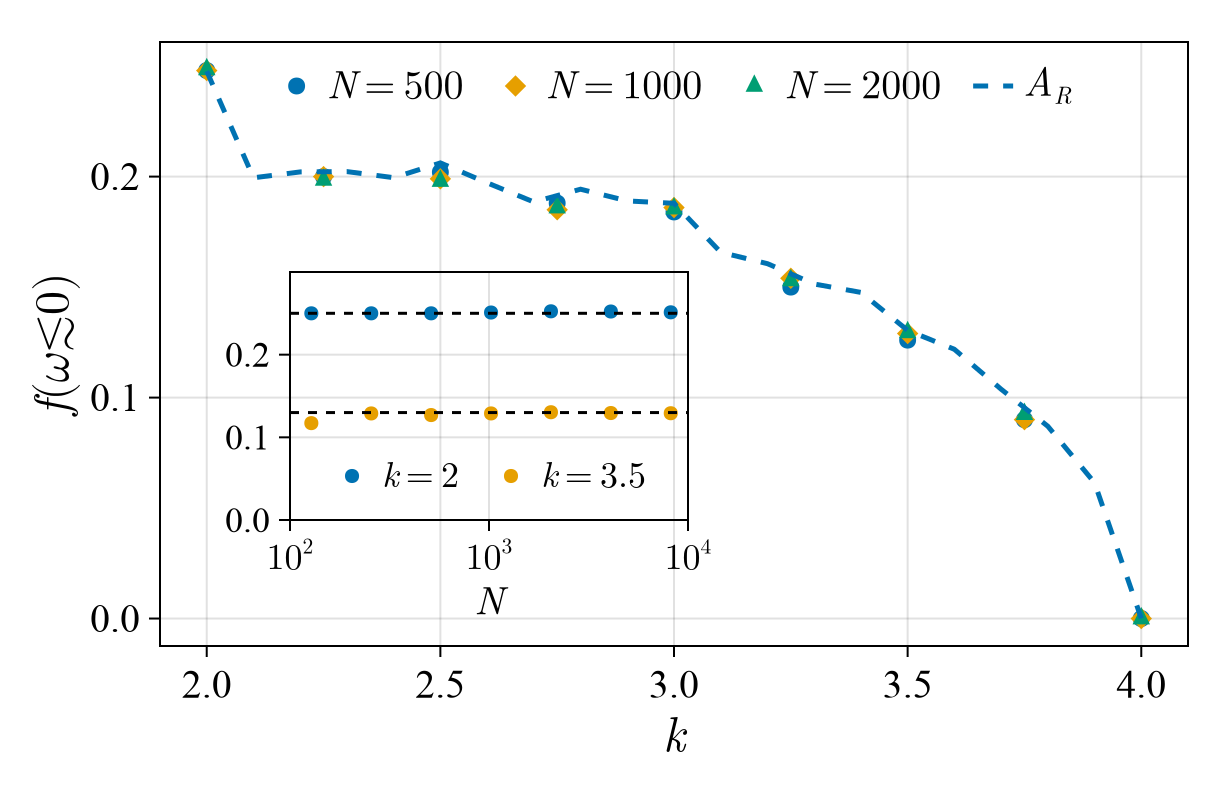}
    \caption{Comparison of the fraction of eigenstates with overlap index $\omega\lesssim 0$ for three different values of $N$, shown as a function of $k$ for map (ii), together with the relative area $A_R$ of the regular island (dashed line). The inset shows the convergence of this fraction with $N$ at two values of $k$. }
    \label{fig:overlap_0}
\end{figure}

In Fig.~\ref{fig:map-joint-dis}, we show the joint distribution for different system sizes, together with example Husimi functions of two representative mixed eigenstates. The distribution exhibits a clear trend in which the overlap index increases with the entropy localization length, except in the limiting regimes $\omega \simeq \pm 1$. This approximately monotonic relation allows us to partition the states equivalently using either measure: the smallest $N_R \simeq N A_R$ states are identified as regular and appear in the lower-left region of the distribution plot, where all regular eigenstates satisfy $\omega \lesssim 0$. This classification is consistent with the Weyl law $N_A \sim (2\pi\hbar)^{-1}\mu(A)$, where $\mu(A)$ is the measure of region $A$ supporting the eigenstates. To further test the validity of $\omega \lesssim 0$ as a criterion for identifying regular eigenstates, Fig.~\ref{fig:overlap_0} compares the fraction of states with $\omega \lesssim 0$ to the relative size of the regular island in map (ii) for different values of $k$ and for various system sizes. The comparison shows consistent convergence across system sizes, confirming that this criterion yields a fraction equal to the relative size of the regular island.

Figure.~\ref{fig:map-joint-dis} also shows that, as the system size increases, the eigenstates cluster more tightly around  $\omega \simeq \pm 1$, and the corresponding localization lengths cluster accordingly. At the same time, more regular states exhibit Husimi functions that partially leak outside the regular region and are therefore identified as mixed eigenstates, the localization length taking close values (for example, those with $-0.8 < \omega \lesssim 0$). In addition, more states lying just outside the regular partition also emerge as mixed eigenstates, exhibiting intermediate localization lengths (for example, those with $0 \lesssim \omega < 0.8$). Although the number of mixed eigenstates increases in both partitions, their overall fraction decays as a power-law with increasing system size as shown in Fig.~\ref{fig:mixed}(d).

\subsection{Sticky eigenstates}
From the preceding analysis of overlap index and localization, the mixed eigenstates can be broadly classified into three typical types:
\begin{itemize}
    \item[(i)] \emph{Regular} states localized on quantized invariant tori near the edge of the main island, and partially leaking out through dynamical tunneling  or due to the finite bump width of the Gaussian coherent state from which the Husimi function is defined, with overlap index $\omega \lesssim 0$.
    
    \item[(ii)] \emph{Chaotic} states which reciprocally  flood into the regular region via the same tunneling mechanism.

    \item[(iii)] \emph{Sticky} states localized near the sticky boundaries formed by MUPOs or quasi-periodic trajectories.
\end{itemize}
As we show next in Sec.~\ref{sec:rmt}, using a random matrix model of dynamical tunneling, the fraction of mixed eigenstates from types (i) and (ii) produced by dynamical tunneling decays much more rapidly, scaling as $\hbar\exp(-b/\hbar)$, similar to the direct regular-to-chaotic tunneling rate \cite{bohigas1993manifestations,tomsovic1994chaos,podolskiy2003semiclassical,backer2008regular} (without considering the improvement from resonance-assisted tunneling), where $b>0$ depends on the relative size of the regular region. Therefore, the regular states, whose leakage is enhanced by the finite-size effect of the coherent state, together with the sticky states, dominate the power-law decay of the mixed eigenstate fraction.

In mixed KAM systems, the type (iii) states are replaced by the hierarchical eigenstates:
\begin{itemize}
\item[(iii)] \emph{Hierarchical} states localized near the resonance- and cantori-induced structures at the boundary between regular and chaotic regions.
\end{itemize}
Reference.~\cite{ketzmerick2000new} proposed that the fraction of hierarchical states scales as $f_{hier}\sim \hbar^{\alpha}$ with $\alpha=1-1/\gamma$, where $\gamma$ is the exponent characterizing the cumulative RTD  $Q(t) \sim t^{-\gamma}$ from classical transport. This exponent exhibits a universal character in generic KAM systems: for bounded phase spaces it typically lies in the range $1.5 \le \gamma \le 3$, whereas in unbounded phase spaces one finds $\gamma \simeq 1.5$. It is therefore of interest to test whether this relation also holds in mixed systems of non-KAM type, that is, whether there exists a direct connection between the decay of the fraction of sticky eigenstates and the cumulative RTD exponent characterizing classical stickiness. 

\begin{figure}
    \centering
    \includegraphics[width=1.0\linewidth]{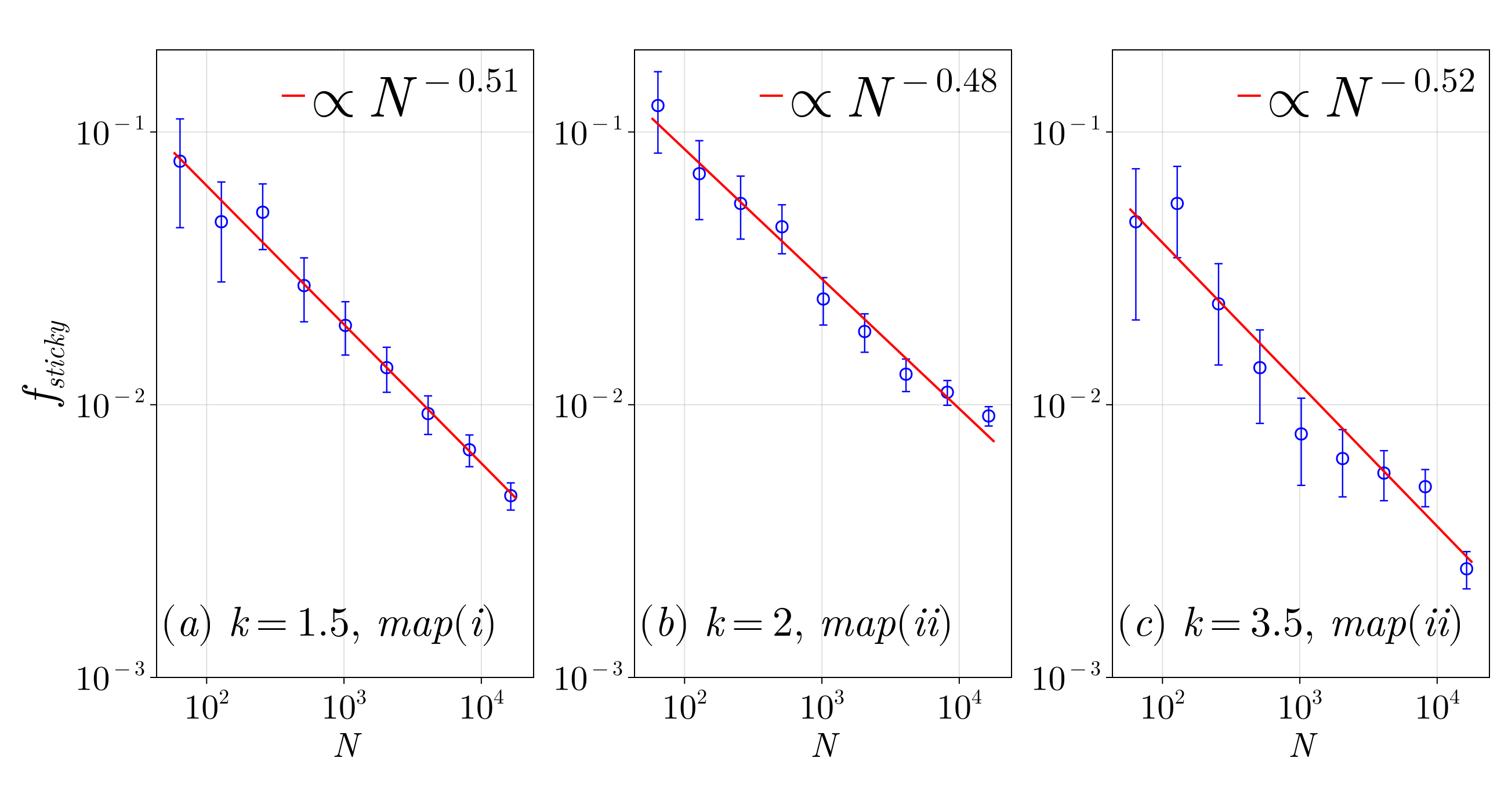}
    \caption{Fraction of sticky eigenstates $f_{sticky}$ with $0\lesssim \omega\le 0.8$ as a function of $N =(2\pi\hbar)^{-1}$, for two maps at different values of $k$. Fits of the fractions are shown as solid lines.}
    \label{fig:sticky}
\end{figure}

Since the regular states of type (i) and the sticky states dominate the mixed eigenstate fraction, the sticky states can be operationally defined by separating out the regular states from the mixed ones, yielding the criterion $0 \lesssim \omega \le \omega_*$. Figure.~\ref{fig:sticky} shows the fraction of sticky eigenstates obtained with $\omega_* = 0.8$ for two maps at different values of $k$. In the first two cases, the fraction $f_{\mathrm{sticky}}$ exhibits a power-law scaling $f_{\mathrm{sticky}} \sim \hbar^{\alpha}$ with an exponent $\alpha$ close to $1 - 1/\gamma$, consistent with the classical transport result that the cumulative RTDs near the MUPOs $Q(t) \sim t^{-\gamma}$ with $\gamma = 2$. In the third case, however, $f_{\mathrm{sticky}}$ does not follow an exact power-law and instead oscillates around the expected scaling $\hbar^{1-1/\gamma}$ with $\gamma = 2$, while the corresponding classical RTD likewise oscillates around the nominal behavior $t^{-2}$ (see Appendix~\ref{app:rts}).

\section{Random matrix model}
\label{sec:rmt}
The Hamiltonian of a mixed system can be written in the following form \cite{bohigas1993manifestations,tomsovic1994chaos,podolskiy2003semiclassical}
\begin{align}
    H=&\sum_R E_R|\psi_R\rangle\langle \psi_R| +\sum_C E_C|\psi_C\rangle\langle\psi_C|\nonumber\\ &+ \sum_{RC}\{ V_{RC}|\psi_R\rangle\langle \psi_C| + c.c.\},
\end{align}
where $E_R$ and $\psi_R$ are the energies and wave functions of the regular states, $E_C$ and $\psi_C$ are the energies and wave functions of the chaotic states, $V_{RC}$ accounts for the interaction between the regular and chaotic basis states. A random matrix model for this Hamiltonian is in the following block structure:
\begin{equation}
H =
\left(
\begin{array}{c|c}
H_{R} & V \\[2pt]
\hline
\\[-3pt]
V^{\dagger} & H_{C}
\end{array}
\right),
\end{equation}
where $H_R$ and $H_C$ describe the regular and chaotic subspaces, respectively, and $V$ denotes the couplings. 

Random matrix models have been successfully applied to mixed systems to describe level splittings produced by chaos-assisted tunneling. In our case, we can adopt the same framework to assess how dynamical tunneling influences the mixed eigenstates. The first block $H_R$ of dimension $N_R$, which models the $N_R$ regular basis states that reside in the regular region of the phase space. For the quantum maps considered here, the phase space contains a single regular island embedded in a chaotic sea. Accordingly, $H_R$ is taken to be a diagonal matrix with eigenvalues $E_R =(k+\xi)/N_R$, $k = 0,1,\ldots, N_R-1$, where the parameter $\xi$ is drawn from a uniform distribution on $[0,1]$. The chaotic block $H_C$ of dimension $N_C$ is from the Gaussian orthogonal ensemble (GOE) with energies $E_C$ in the interval $[-1,1]$, or from circular orthogonal ensemble (COE). The diagonal structure of the regular block $H_R$ and the random matrix structure of the chaotic block $H_C$ encode, respectively, the localization of the regular eigenstates $\psi_R$ and the delocalization of the chaotic eigenstates $\psi_C$ on the computational basis.

\begin{figure}
    \centering
    \includegraphics[width=1.0\linewidth]{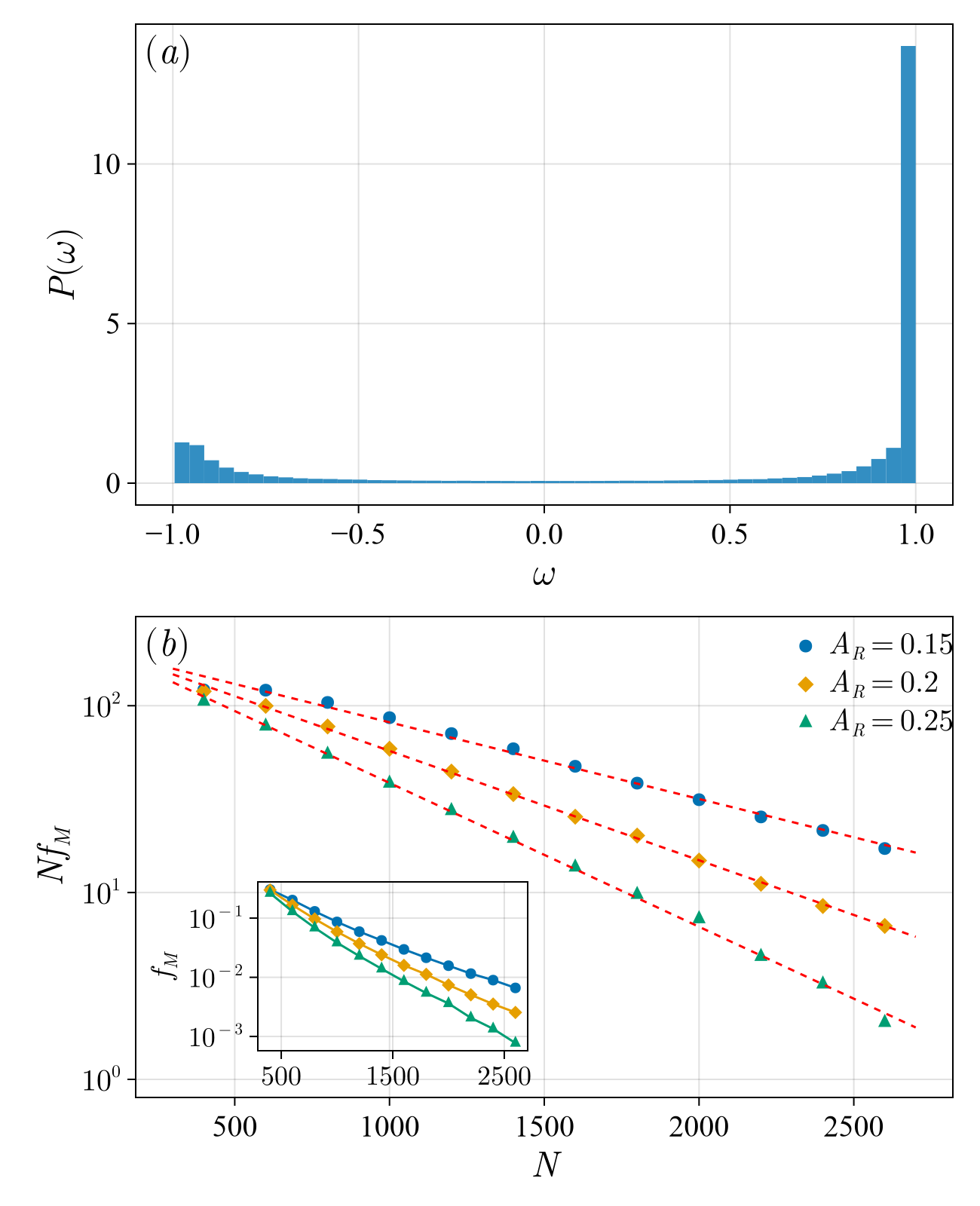}
    \caption{(a) Histogram of the overlap index $\omega$ for $A_R=0.25$ and  $N=500$. (b) Rescaled fraction of mixed eigenstates $f_M$ with $|\omega|\le 0.8$ as a function of $N$ from the random matrix model, for three values of $A_R$. Dotted lines indicate fits to the rescaled data, and the insets display the corresponding fractions. Each data point is obtained from an ensemble of $\sim 4\times 10^5$ eigenvectors at each $N$.}
    \label{fig:rmt1}
\end{figure}

The off-diagonal block $V$ accounts for the coupling between regular and chaotic states. Its matrix elements $V_{RC}$ fluctuate strongly and are modeled as independent Gaussian random numbers with variance \cite{podolskiy2003semiclassical}
\begin{align}
\label{eq:coupling}
    \langle |V_{RC}|^2\rangle = c_0 \hbar_{\text{eff}}^2\frac{\Gamma(\frac{A_R}{\pi\hbar_{\text{eff}}},\frac{2A_R}{\pi\hbar_{\text{eff}}})}{\Gamma(\frac{A_R}{\pi\hbar_{\text{eff}}}+1,0)},
\end{align}
where $A_R=N_R/(N_R+N_C)=N_R/N$ is the relative size of the regular island in phase space, $\hbar_{\text{eff}}=(2\pi N)^{-1}$ is the effective Planck constant, $\Gamma$ denotes the incomplete Gamma function and $c_0$ is a nonuniversal prefactor that does not depend on $\hbar_{\text{eff}}$ in random matrix models and in the following, we set $c_0 = 1$. We define the overlap index of the eigenvectors $\psi_n$ of $H$ in the random matrix model 
\begin{align}
\label{eq:rmt-index}
    \omega_n = \sum_{i=0}^{N-1} \chi_i |\langle i|\psi_n\rangle|^2,
\end{align}
where $\{|i\rangle\}$ denotes the computational basis. The basis is ordered such that the states $\{|i\rangle : 0 \le i \le N_R - 1\}$ correspond to  the $i$th quantized torus in the regular block $H_R$, while the states $\{|i\rangle : N_R \le i \le N - 1\}$  correspond to the chaotic block $H_C$, on which the eigenstates of $H_C$ are delocalized. The characteristic function $\chi_i$ takes values on the computational basis as
\begin{equation}
    \chi_i =
    \begin{cases}
        -1,&  0 \le i \le N_R - 1, \\
        +1,&  N_R \le i \le N - 1,
    \end{cases}
\end{equation}
in direct analogy with Eq.~\eqref{eq:olap-map} in physical systems.

If we set the coupling term $V_{RC}$ to zero, the model obviously realizes the PUSC scenario: regular eigenstates have an overlap index $\omega = -1$, while chaotic eigenstates have $\omega = +1$ and are fully delocalized as random vectors. In this limit, the level spacing distribution follows the Berry-Robnik statistics. Introducing the coupling defined in Eq.~\eqref{eq:coupling} gives rise to direct regular-to-chaotic tunneling (i.e., dynamical tunneling), which leads to fractional level repulsion at small spacings, with the tunneling rates scaling as $\gamma_T \sim \hbar_{\mathrm{eff}} \exp(-a/\hbar_{\mathrm{eff}})$ from Fermi's golden rule, where the coefficient $a>0$ is associated with the relative size of the regular region $A_R$ and larger regular regions yield larger values of $a$, consistent with the behavior observed in two-dimensional maps \cite{backer2008regular}.

This tunneling mechanism also yields intermediate values of the overlap index $|\omega|<1$, as seen in Fig.~\ref{fig:rmt1}(a), which shows the corresponding histogram of the overlap index for the random matrix model with $A_R = 0.25$ and $N = 500$. Furthermore, Fig.~\ref{fig:rmt1}(b) demonstrates that the fraction of mixed eigenstates decays with the effective Planck constant in the same functional form as the regular-to-chaotic tunneling rate 
\begin{align}
    f_M \sim N^{-1}\exp(-bN) \sim  \hbar_{\mathrm{eff}} \exp(-b/\hbar_{\mathrm{eff}}),
\end{align}
where $b>0$ is another coefficient that increases with the size of the regular region $A_R$. As shown in Fig.~\ref{fig:rmt1}, larger $A_R$ produces a correspondingly steeper slope in the log-normal plot. Here we set $\omega^*=0.8$, and similar results are obtained for other values of $\omega^* <1$. 

The entropy localization length of mixed eigenstates can also be characterized in the random matrix model,  through the Shannon entropy of the eigenvectors
\begin{align}
    S_n=-\sum_{i=0}^{N-1}|\psi_{ni}|^2\ln |\psi_{ni}|^2,
\end{align}
with $l_n=\exp(S_n)/N$, where $\psi_{ni}\equiv \langle i|\psi_n\rangle$. For a random pure state from the COE or GOE, the expansion components $\psi_{ni}$ are all real random numbers, the mean Shannon entropy \cite{izrailev1988quantum}
\begin{align}
    \langle S\rangle_N \sim \ln N + \gamma_e+\ln 2-2.
\end{align}
For $N\gg 1$, the maximum entropy localization length of the eigenvectors can be approximated, given by the mean value of random pure states, is $l_{\max} \simeq e^{\gamma_e+\ln2-2} \simeq 0.482$. 

\begin{figure}
    \centering
    \includegraphics[width=1.0\linewidth]{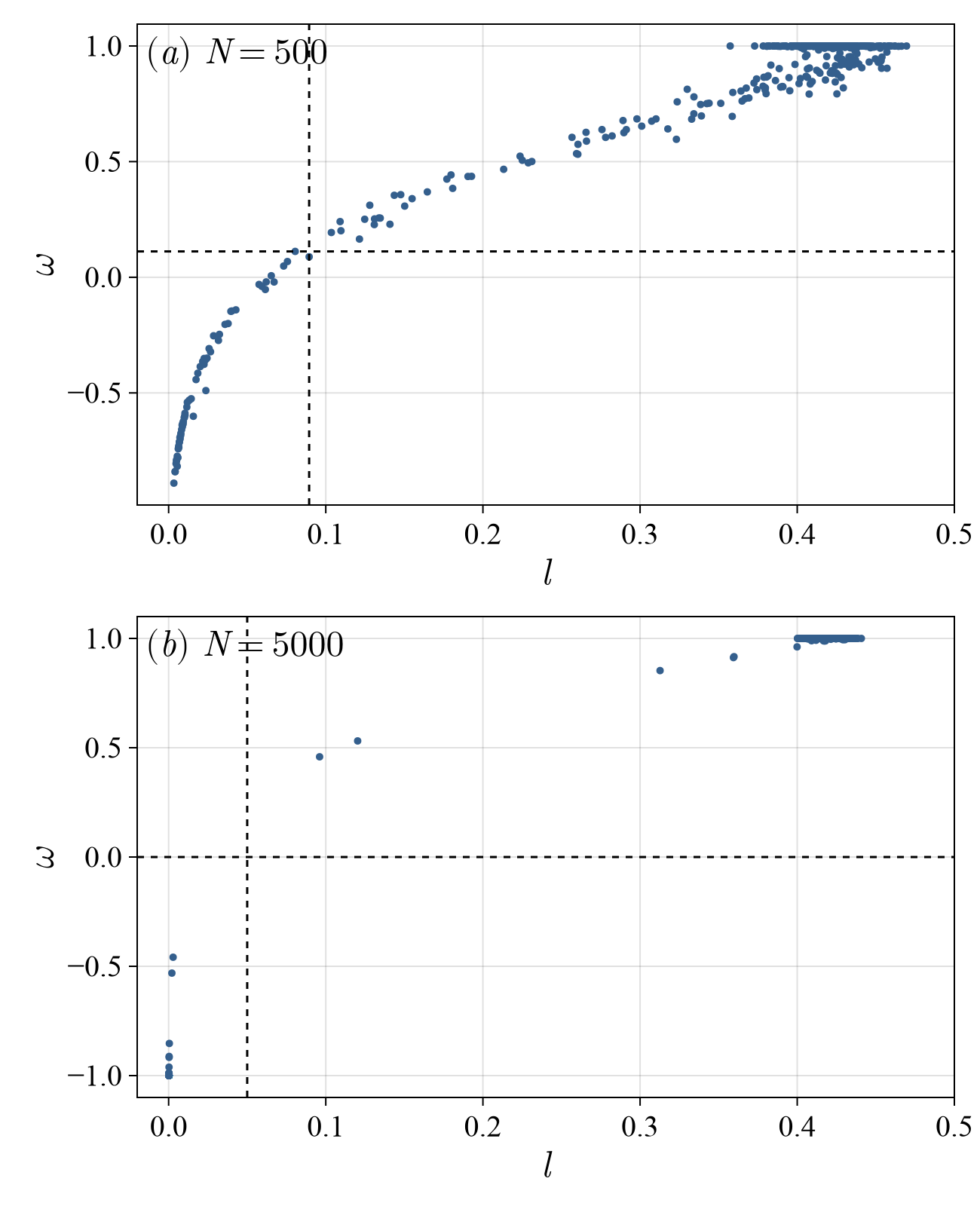}
    \caption{Analogous to Fig.~\ref{fig:map-joint-dis}, the joint distribution of entropy localization length $l$ and the overlap index $\omega$ for the random matrix model with the same $A_R=0.13$. The vertical and horizontal dashed lines partition the states as Fig.~\ref{fig:map-joint-dis} into the lower-left and upper-right groups, where the number of states in the lower-left group is $NA_R$.}
    \label{fig:rm-joint-dis}
\end{figure}

In Fig.~\ref{fig:rm-joint-dis}, we show the joint distribution for two values of $N$ to compare with the quantum maps discussed above and the results in Fig.~\ref{fig:map-joint-dis}. The correlation between the entropy localization length and the overlap index is evident: the overlap index grows with the localization length, except in the limiting regimes $\omega \simeq \pm 1$. As the system size increases, the eigenstates cluster more tightly around $\omega \simeq \pm 1$, and the corresponding localization lengths condense accordingly, consistent with the behavior of the quantum maps. In contrast, a smaller number of regular states tunnel out of the regular region and become mixed, and likewise fewer states outside the regular partition appear as mixed eigenstates. As a result, their fraction decays as $\hbar \exp(-b/\hbar_{\mathrm{eff}})$, much faster than the algebraic decay observed in the quantum maps, implying that mixed eigenstates generated by dynamical tunneling can be neglected in maps. This discrepancy arises for two reasons intrinsic to the random matrix model: first, the definition of the overlap index in Eq.~\eqref{eq:rmt-index} lacks the finite-size smearing associated with coherent states in physical systems, since it is constructed upon the computational basis and therefore does not identify regular states as mixed; and second, the model contains no mechanism for stickiness, and therefore no sticky eigenstates.

\section{Summary and Conclusions}
\label{sec:sum}
We investigate mixed eigenstates in non-KAM systems with sharply divided phase space, specifically two different piecewise-linear maps, in which the boundary between the regular island and the chaotic sea is formed by MUPOs or quasi-periodic orbits and where classical stickiness is present. For MUPOs, the cumulative RTD follows $Q(t) \sim t^{-\gamma}$ with $\gamma = 2$, while in the quasi-periodic case $Q(t)$ oscillates around $t^{-2}$. We confirm that the fraction of mixed eigenstates  $f_M$ exhibits a power-law decay, scaling as $f_M\sim \hbar^{\eta}$ with $\eta>0$, as in KAM systems. Using the overlap index and the entropy localization length, we classify the mixed eigenstates into three types: regular states that arise from finite-size of the Gaussian bump of the coherent state underlying the Husimi function or that leak into the chaotic region through dynamical tunneling, chaotic states that reciprocally flood the regular region through the same tunneling mechanism, and sticky eigenstates localized near the sticky boundaries. 

From random matrix modeling, we show that the fraction of states from dynamical tunneling scales as $\hbar \exp(-b/\hbar)$, with $b>0$ associated with the relative size of the regular block/region. Consequently, the observed power-law decay of mixed eigenstates is dominated by regular states that are classified as mixed because of the finite size of the coherent state, as well as by the sticky eigenstates. We further demonstrate that regular states identified as mixed eigenstates can be effectively separated using the criterion $\omega \lesssim 0$.  This allows us to determine the fraction of sticky eigenstates $f_{\text{sticky}}$. We find that $f_{\text{sticky}}$ scales as $\hbar^{1 - 1/\gamma}$ with $\gamma = 2$ for maps in which MUPOs form the sticky boundary.  For map with an elliptical regular region, where a quasi-periodic orbit serves as the sticky boundary, $f_{\text{sticky}}$ oscillates around $t^{-2}$. This scaling of the sticky eigenstates is consistent with the proposal in KAM systems concerning the scaling behavior of hierarchical eigenstates that originate from the hierarchical structure of the boundary.  

\begin{figure*}
    \centering
    \includegraphics[width=0.9\linewidth]{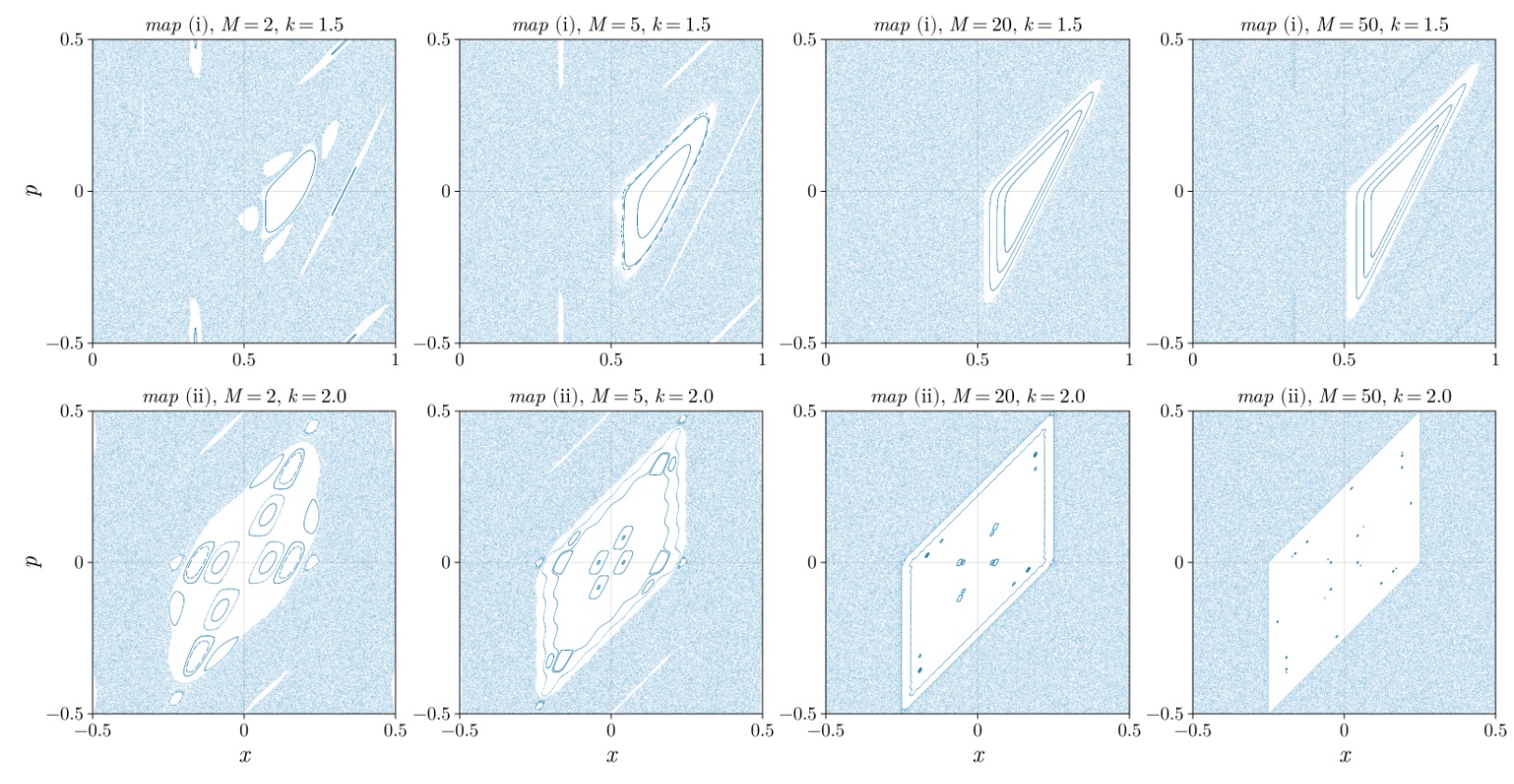}
    \caption{Phase-space portraits for map (i) at $k=1.5$ (upper panels) and map (ii) at $k=2.0$ (lower panels), generated using the smoothed potentials in Eqs.~\eqref{eq:map1-smooth} and \eqref{eq:map2-smooth}. Each portrait is formed from 30 random trajectories evolved for $1\times 10^4$ iterations. From left to right, the panels show $M=2,5,20,50$.}
    \label{fig:limit-ps}
\end{figure*}

We note that the criterion used in the original study to separate hierarchical eigenstates, based on Berry–Robnik statistics together with the assumption that chaotic eigenstates follow the Gaussian random-wave form across all system sizes,  does not take into account the effects of dynamical localization in chaotic eigenstates or the associated spectral statistics. These omissions lead to noticeable fluctuations in numerical results. The numerical analysis in KAM systems therefore can be refined by introducing the overlap index and applying the criterion developed in this work.  It is also worth mentioning that sticky eigenstates contribute to resonance states with small decay rates when the system is opened via the leaking formalism. {\note As shown in Ref.~\cite{ishii2012weyl}, their fraction in a two-dimensional piecewise-linear map follows $\hbar^{1/2}$, where the boundary is formed by MUPOs, consistent with the behavior of $1-1/\gamma_e$, where $\gamma_e=2$ denotes the exponent of the power-law escape, or survival probability. It may therefore be viewed as an extension of the fractal Weyl law verified in open chaotic systems \cite{lu2003fractal}. This suggests that the equivalence between the recurrence time distribution and an escape time distribution obtained by leaking the system \cite{altmann2009poincare}, as well as the connection between sticky eigenstates in the unitary system and their fate in the corresponding open system, an interesting direction for future research.}

An open question arising from this work is whether an effective theory can explain the apparent closeness between the power-law exponent of the sticky eigenstates and the power-law exponent characterizing classical stickiness in non-KAM system, where no hierarchical structures exist near the sticky boundary and the self-similar Markov tree model is therefore not applicable. It would also be interesting to investigate the topological properties of these sticky or hierarchical eigenstates.

\section{Acknowledgements}
The author thanks M. Robnik for his inspiring ideas and careful reading of the manuscript, A. Shudo and H. Skokos for clarifying discussions. This work was supported by the Slovenian Research and Innovation Agency (ARIS) under grants J1-4387 and P1-0306, and by the HPC Vega infrastructure. The author used ChatGPT for assistance with grammar and proofreading.

\appendix
\section{Disappearing of tiny islands}
\label{app:island}

{\note 
In Sec.~\ref{sec:sec2a}, we construct a smooth approximation \(f_M(x)\) to the piecewise-linear function \(f(x)\) appearing in the two maps by truncating the Taylor expansion. This approximation converges to the original piecewise-linear form in the limit \(M\to\infty\), namely \(f(x)=\lim_{M\to\infty} f_M(x)\). For small \(M\), the smoothing introduced by \(f_M(x)\) gives rise to tiny island chains surrounding the main regular island, consistent with the KAM theorem. As \(M\) increases, these island chains become less visible as the border between the regular and chaotic regions becomes sharper, approaching the border formed by MUPOs in the piecewise-linear case shown in Fig.~\ref{fig:cp1}. This behavior is illustrated in Fig.~\ref{fig:limit-ps}, by the phase-space portraits for \(M=2,5,20,\) and \(50\), where for $M=50$ almost no secondary island chains or other tiny islands are visible around the main regular island, indicating the general tendency of island chains to disappear in the approach to the piecewise-linear limit (a $C^2$ discontinuity), where the smoothness assumptions of the KAM theorem cease to hold. }

\section{Recurrence-time distributions}
\label{app:rts}
 Stickiness in Hamiltonian systems is quantified by the recurrence-time distribution $P(t)$, which reveals the algebraic trapping behavior near regular structures. The recurrence-time measures the interval required for a trajectory to return to a predefined region away from the regular island. In practice, it is often more convenient to analyze its cumulative distribution
\begin{align}
Q(t) = \sum_{t' = t}^{\infty} P(t') = \lim_{N\to\infty}\frac{N_t}{N},
\end{align}
gives the probability of remaining trapped longer than $t$, where $N_t$ is the number of recurrences with time $t'\ge t$ and $N$ is the total number of recurrence events observed in the long time limit. For \emph{fully} chaotic systems, $Q(t)$ decays exponentially at long times, $Q(t) \sim \exp(-h t)$, where $h$ is the Kolmogorov-Sinai entropy. In mixed KAM systems, cantori-induced stickiness yields a power-law decay of $Q(t)$ instead. The corresponding exponent displays a universal character: for bounded phase spaces it typically satisfies $1.5 \le \gamma \le 3$, whereas in unbounded phase spaces one finds $\gamma \simeq 1.5$ \cite{cristadoro2008universality,venegeroles2009universality}.

\begin{figure}
    \centering
    \includegraphics[width=1.0\linewidth]{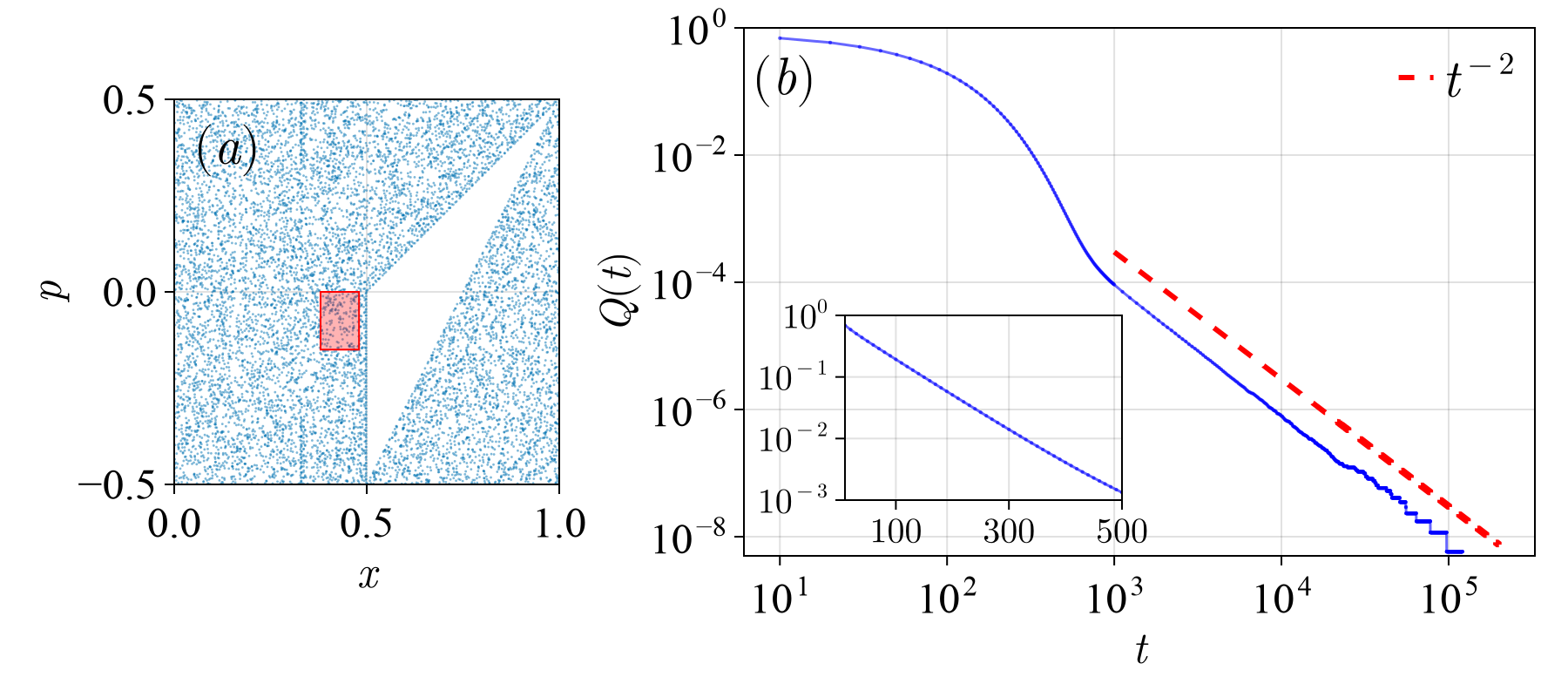}
    \vspace{0.2cm}
    \includegraphics[width=1.0\linewidth]{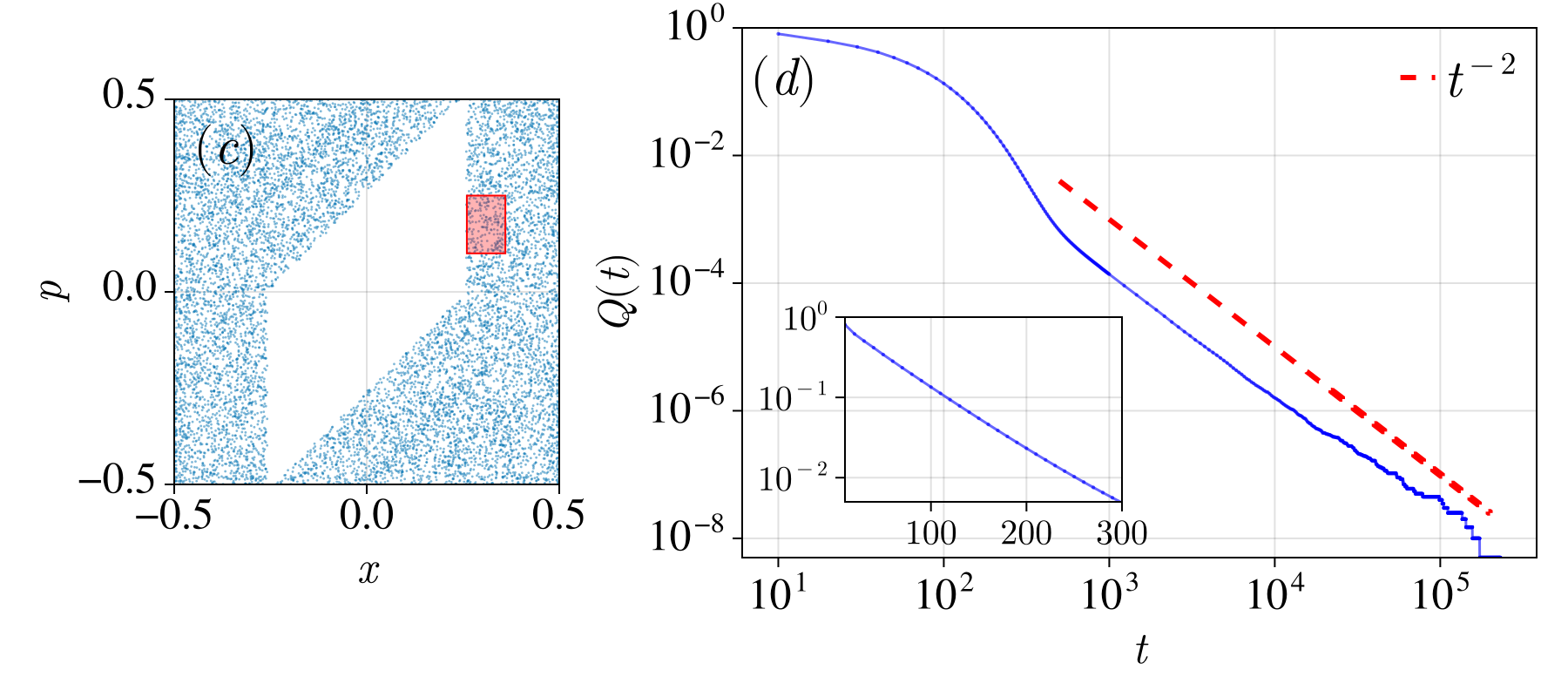}
    \caption{(a) Rectangular recurrence region (in red) near the regular island used for the recurrence-time distributions in map (i) at $k=1.5$. (b) Cumulative RTDs, where the inset shows the same plot for short time regime in the log-normal scale. The distribution is an exponential distribution for the short time regime and a power-law distribution $\sim t^{-2}$ for the long-time regime. (c–d) Same as panels (a–b), but for map (ii) at $k=2$.}
    \label{fig:rtd}
\end{figure}

In mixed systems of non-KAM type, including mushroom billiards  and piecewise-linear maps with sharply divided phase space where MUPOs form the sticky boundaries, it has been shown that $\gamma=2$ \cite{bunimovich2001mushrooms,altmann2005stickiness,barnett2007quantum}. A more detailed analysis further indicates that, due to the intrinsic inhomogeneity of the phase space, generic mixed systems exhibit finite-size effects associated with the chosen recurrence region: the RTD may be viewed as a superposition of two contributions: an exponential component reflecting the hyperbolic nature of the dynamics, and a power-law component arising from sticky non-hyperbolic structures \cite{akaishi2009accumulation,altmann2009poincare}. Accordingly,
\begin{align}
    Q(t)\sim
\begin{cases}
C_1 e^{-\mu t}, & t < t_p, \\[6pt]
C_1 e^{-\mu t} + C_2 t^{-2}, & t \ge t_p ,
\end{cases}
\end{align}
where $C_1,C_2$ are normalization constant, $\mu$ is determined by the measure of the recurrence region and $t_p$ marks the onset of the power-law regime. This form captures the exponential distribution at short time regime and $t^{-2}$ power-law that dominates in the long-time, as clearly shown in Fig.~\ref{fig:rtd} for two classical maps whose regular–chaotic boundaries are formed by MUPOs.

\begin{figure}
    \centering
    \includegraphics[width=1\linewidth]{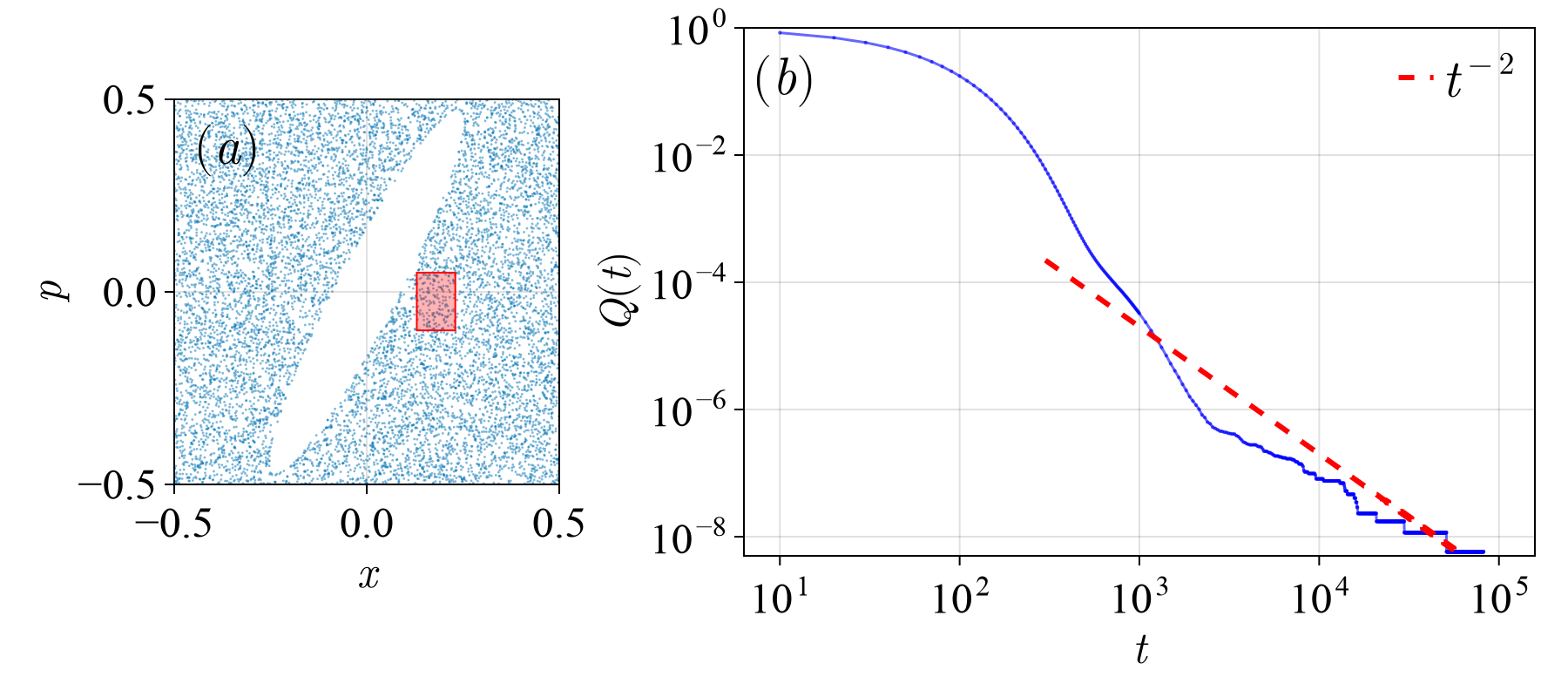}
    \caption{Analogous to Fig.~\ref{fig:rtd} but for map (ii) at $k=3.5$. In this case, the boundary are formed by a quasi-periodic trajectory instead of MUPOs. Oscillations arise in the long-time regime, while the results deviate from the $t^{-2}$ decay, and it is the \emph{average} trend that remains consistent with it.}
    \label{fig:rtd-quasi}
\end{figure}

For map (ii) at $k = 3.5$, or at other parameter values that produce a sharply divided phase space with an elliptically shaped regular island, the boundary between the island and the chaotic component is no longer formed by MUPOs, instead, it is formed by a quasi-periodic trajectory. The RTD does not follow the power-law $t^{-2}$ exactly at long times, but instead exhibits oscillations around it, as shown in Fig.~\ref{fig:rtd-quasi}.

\section{Approximation of the coherent state on torus}
\label{app:asymp}
Eq.~\eqref{eq:cs-torus} gives the expression of the coherent state on the torus, in the form of a Jacobi-Theta function which is invariant under translation in both the position and momentum. The evaluation of these coherent states can be simplified first by combining all the real parts and imaginary parts term by term
\begin{align}
    \mathcal{E}&=e^{-2\pi N[(\bar{\alpha}^2+|\alpha|^2+x_j^2)/2-\sqrt{2}\bar{\alpha}x_j]} 
    \theta_3\left(i N[x_j-\sqrt{2}\bar{\alpha}]|iN\right)\nonumber\\
    &=\sum_{\nu\in\mathbb{Z}}\exp\left(-\pi N(\nu+x_j-x)^2\right)\nonumber\\
    &\qquad\qquad\qquad\times\exp\left(-i\pi Np(2x_j-x+2\nu)\right).
\end{align}
Let $\Delta=x-x_j\in[-1,1]$, and define $r=\Delta-\lfloor\Delta+\frac{1}{2}\rfloor$. Equivalently, 
$|r|=dist(\Delta,\mathbb{Z})$ is the distance from $\Delta$ to the nearest integer. Clearly, $|r|\le 1/2$. For $N\gg 1$ the sum is dominated by a single Gaussian term
\begin{align}
    \mathcal{E}\simeq \mathcal{E}_0=\exp(-\pi Nr^2)\exp\left(-i\pi Np(x+2r)\right),
\end{align}
where a \emph{convenient} upper bound on the truncation error
\begin{align}
    |\mathcal{E}-\mathcal{E}_0|\le \sum_{m\ge 1} 2e^{-\pi N(r+m)^2}\le \frac{2e^{-\pi N(1-|r|)^2}}{1-e^{-\pi N(2-2|r|)}},
\end{align}
this upper bound goes exponentially small for $N\gg 1$, so the single Gaussian term $\mathcal{E}_0$ is extremely accurate.

% \section{Additional results for elliptical regular region cases}

% \begin{figure}
%     \centering
%     \includegraphics[width=1.0\linewidth]{more_loc_olap.png}
%     \caption{}
%     \label{fig:placeholder}
% \end{figure}

% \begin{figure}
%     \centering
%     \includegraphics[width=1.0\linewidth]{more_ellipse_decay.png}
%     \caption{Caption}
%     \label{fig:placeholder}
% \end{figure}

\bibliographystyle{apsrev4-2}
\bibliography{ref}

@article{robnik1998topics,
	author = {Robnik, M},
	journal = {Nonlinear Phenomena in Complex Systems},
	number = {1-4},
	pages = {1--22},
	title = {Topics in Quantum Chaos in Generic Systems.},
	volume = {1},
	year = {1998}}

@article{backer2008dynamical,
  title={Dynamical tunneling in mushroom billiards},
  author={B{\"a}cker, A and Ketzmerick, R and L{\"o}ck, S and Robnik, Marko and Vidmar, Gregor and H{\"o}hmann, Ruven and Kuhl, Ulrich and St{\"o}ckmann, H-J},
  journal={Physical review letters},
  volume={100},
  number={17},
  pages={174103},
  year={2008},
  publisher={APS}
}

@article{brun1999realizing,
  title={Realizing the quantum baker’s map on a NMR quantum computer},
  author={Brun, Todd A and Schack, R{\"u}diger},
  journal={Physical Review A},
  volume={59},
  number={4},
  pages={2649},
  year={1999},
  publisher={APS}
}

@article{lu2003fractal,
  title={Fractal Weyl laws for chaotic open systems},
  author={Lu, WT and Sridhar, Srinivas and Zworski, Maciej},
  journal={Physical review letters},
  volume={91},
  number={15},
  pages={154101},
  year={2003},
  publisher={APS}
}

@article{altmann2005stickiness,
  title={Stickiness in mushroom billiards},
  author={Altmann, Eduardo G and Motter, Adilson E and Kantz, Holger},
  journal={Chaos: An Interdisciplinary Journal of Nonlinear Science},
  volume={15},
  number={3},
  year={2005},
  publisher={AIP Publishing}
}

@article{robnik2023recent,
  title={Recent Developments in Quantum Chaos of Mixed-Type Systems: A Mini Review},
  author={Robnik, Marko},
  journal={Nonlinear Phenomena in Complex Systems},
  volume={26},
  number={3},
  pages={209--224},
  year={2023}
}

@article{batistic2013dynamical,
  title={Dynamical localization of chaotic eigenstates in the mixed-type systems: spectral statistics in a billiard system after separation of regular and chaotic eigenstates},
  author={Batisti{\'c}, Benjamin and Robnik, Marko},
  journal={Journal of Physics A: Mathematical and Theoretical},
  volume={46},
  number={31},
  pages={315102},
  year={2013},
  publisher={IOP Publishing}
}

@article{altmann2009poincare,
  title={Poincar{\'e} recurrences and transient chaos in systems with leaks},
  author={Altmann, Eduardo G and T{\'e}l, Tam{\'a}s},
  journal={Physical Review E—Statistical, Nonlinear, and Soft Matter Physics},
  volume={79},
  number={1},
  pages={016204},
  year={2009},
  publisher={APS}
}

@article{ketzmerick1996fractal,
  title={Fractal conductance fluctuations in generic chaotic cavities},
  author={Ketzmerick, Roland},
  journal={Physical Review B},
  volume={54},
  number={15},
  pages={10841},
  year={1996},
  publisher={APS}
}

@article{alhassid2000statistical,
  title={The statistical theory of quantum dots},
  author={Alhassid, Yoram},
  journal={Reviews of Modern Physics},
  volume={72},
  number={4},
  pages={895},
  year={2000},
  publisher={APS}  
}

@article{huckestein2000quantum,
  title={Quantum transport through ballistic cavities: Soft vs hard quantum chaos},
  author={Huckestein, Bodo and Ketzmerick, Roland and Lewenkopf, Caio H},
  journal={Physical Review Letters},
  volume={84},
  number={24},
  pages={5504},
  year={2000},
  publisher={APS}
}

@article{dettmann2011open,
  title={Open mushrooms: stickiness revisited},
  author={Dettmann, Carl P and Georgiou, Orestis},
  journal={Journal of Physics A: Mathematical and Theoretical},
  volume={44},
  number={19},
  pages={195102},
  year={2011}
}

@article{serbyn2021quantum,
  title={Quantum many-body scars and weak breaking of ergodicity},
  author={Serbyn, Maksym and Abanin, Dmitry A and Papi{\'c}, Zlatko},
  journal={Nature Physics},
  volume={17},
  number={6},
  pages={675--685},
  year={2021},
  publisher={Nature Publishing Group UK London}
}

@article{orel2026quantum,
  title={Quantum chaos and semiclassical behavior in mushroom billiards I: Spectral statistics},
  author={Orel, Matic and Lozej, {\v{C}}rt and Robnik, Marko and Yan, Hua},
  journal={Physical Review E},
  volume={113},
  number={4},
  pages={044217},
  year={2026},
  publisher={APS}
}

@article{moudgalya2022quantum,
  title={Quantum many-body scars and Hilbert space fragmentation: A review of exact results},
  author={Moudgalya, Sanjay and Bernevig, B Andrei and Regnault, Nicolas},
  journal={Reports on Progress in Physics},
  volume={85},
  number={8},
  pages={086501},
  year={2022},
  publisher={IOP Publishing}
}

@article{weinstein2002experimental,
  title={Experimental implementation of the quantum baker’s map},
  author={Weinstein, Yaakov S and Lloyd, Seth and Emerson, Joseph and Cory, David G},
  journal={Physical review letters},
  volume={89},
  number={15},
  pages={157902},
  year={2002},
  publisher={APS}
}

@article{patoary2024chaotic,
  title={Chaotic roots of the modular multiplication dynamical system in Shor's algorithm},
  author={Patoary, Abu Musa and Vikram, Amit and Shou, Laura and Galitski, Victor},
  journal={Physical Review Research},
  volume={6},
  number={3},
  pages={L032046},
  year={2024},
  publisher={APS}
}

@article{benenti2026quantum,
  title={The Quantum Kicked Rotor: A Paradigm of Quantum Chaos. Foundational aspects and new perspectives},
  author={Benenti, Giuliano and Casati, Giulio and Gong, Jiangbin and Zou, Zhixing},
  journal={arXiv preprint arXiv:2604.12316},
  year={2026}
}

@article{altmann2013leaking,
  title={Leaking chaotic systems},
  author={Altmann, Eduardo G and Portela, Jefferson SE and T{\'e}l, Tam{\'a}s},
  journal={Reviews of Modern Physics},
  volume={85},
  number={2},
  pages={869--918},
  year={2013},
  publisher={APS}
}

@book{keshavamurthy2011dynamical,
  title={Dynamical tunneling: theory and experiment},
  author={Keshavamurthy, Srihari and Schlagheck, Peter},
  year={2011},
  publisher={CRC Press}
}

@article{yan2025spacing,
  title={Spacing ratios in mixed-type systems},
  author={Yan, Hua},
  journal={Physical Review E},
  volume={111},
  number={5},
  pages={054213},
  year={2025},
  publisher={APS}
}

@article{barnett2007quantum,
  title={Quantum mushroom billiards},
  author={Barnett, Alex H and Betcke, Timo},
  journal={Chaos: An Interdisciplinary Journal of Nonlinear Science},
  volume={17},
  number={4},
  year={2007},
  publisher={AIP Publishing}
}

@article{bunimovich2001mushrooms,
	author = {Bunimovich, Leonid A},
	journal = {Chaos: An Interdisciplinary Journal of Nonlinear Science},
	number = {4},
	pages = {802--808},
	publisher = {American Institute of Physics},
	title = {Mushrooms and other billiards with divided phase space},
	volume = {11},
	year = {2001}}

@article{izrailev1988quantum,
	author = {Izrailev, FM},
	journal = {Physics letters A},
	number = {1},
	pages = {13--18},
	publisher = {Elsevier},
	title = {Quantum localization and statistics of quasienergy spectrum in a classically chaotic system},
	volume = {134},
	year = {1988}}

@article{gnutzmann2001renyi,
	author = {Gnutzmann, Sven and Zyczkowski, Karol},
	journal = {Journal of Physics A: Mathematical and General},
	number = {47},
	pages = {10123},
	publisher = {IOP Publishing},
	title = {R{\'e}nyi-Wehrl entropies as measures oflocalization in phase space},
	volume = {34},
	year = {2001}}

@article{lieb1978proof,
	author = {Lieb, Elliott H},
	journal = {Communications in Mathematical Physics},
	number = {1},
	pages = {35--41},
	publisher = {Springer},
	title = {Proof of an entropy conjecture of Wehrl},
	volume = {62},
	year = {1978}}

@article{wehrl1979relation,
	author = {Wehrl, Alfred},
	journal = {Reports on Mathematical Physics},
	number = {3},
	pages = {353--358},
	publisher = {Elsevier},
	title = {On the relation between classical and quantum-mechanical entropy},
	volume = {16},
	year = {1979}}

@article{yan2024further,
	author = {Yan, Hua and Wang, Qian and Robnik, Marko},
	journal = {Physical Review E},
	number = {6},
	pages = {064222},
	publisher = {APS},
	title = {Further results on the power-law decay of the fraction of the mixed eigenstates in kicked-top model with mixed-type classical phase space},
	volume = {110},
	year = {2024}}

@article{wang2023power,
	author = {Wang, Qian and Robnik, Marko},
	journal = {Physical Review E},
	number = {5},
	pages = {054217},
	publisher = {APS},
	title = {Power-law decay of the fraction of the mixed eigenstates in kicked top model with mixed-type classical phase space},
	volume = {108},
	year = {2023}}

@article{yan2024chaos,
	author = {Yan, Hua and Robnik, Marko},
	journal = {Physical Review E},
	number = {5},
	pages = {054211},
	publisher = {APS},
	title = {Chaos and quantization of the three-particle generic Fermi-Pasta-Ulam-Tsingou model. II. Phenomenology of quantum eigenstates},
	volume = {109},
	year = {2024}}

@article{lozej2022phenomenology,
	author = {Lozej, {\v{C}}rt and Lukman, Dragan and Robnik, Marko},
	journal = {Physical review E},
	number = {5},
	pages = {054203},
	publisher = {APS},
	title = {Phenomenology of quantum eigenstates in mixed-type systems: Lemon billiards with complex phase space structure},
	volume = {106},
	year = {2022}}

@article{meiss1985markov,
	author = {Meiss, James D and Ott, Edward},
	journal = {Physical review letters},
	number = {25},
	pages = {2741},
	publisher = {APS},
	title = {Markov-tree model of intrinsic transport in Hamiltonian systems},
	volume = {55},
	year = {1985}}

@article{cristadoro2008universality,
	author = {Cristadoro, Giampaolo and Ketzmerick, Roland},
	journal = {Physical review letters},
	number = {18},
	pages = {184101},
	publisher = {APS},
	title = {Universality of algebraic decays in Hamiltonian systems},
	volume = {100},
	year = {2008}}

@article{venegeroles2009universality,
	author = {Venegeroles, Roberto},
	journal = {Physical Review Letters},
	number = {6},
	pages = {064101},
	publisher = {APS},
	title = {Universality of algebraic laws in Hamiltonian systems},
	volume = {102},
	year = {2009}}

@article{brodier2001resonance,
	author = {Brodier, Olivier and Schlagheck, Peter and Ullmo, Denis},
	journal = {Physical Review Letters},
	number = {6},
	pages = {064101},
	publisher = {APS},
	title = {Resonance-assisted tunneling in near-integrable systems},
	volume = {87},
	year = {2001}}

@article{batistic2013quantum,
	author = {Batisti{\'c}, Benjamin and Robnik, Marko},
	journal = {Physical Review E---Statistical, Nonlinear, and Soft Matter Physics},
	number = {5},
	pages = {052913},
	publisher = {APS},
	title = {Quantum localization of chaotic eigenstates and the level spacing distribution},
	volume = {88},
	year = {2013}}

@article{chirikov1988quantum,
	author = {Chirikov, BV and Izrailev, FM and Shepelyansky, DL},
	journal = {Physica D: Nonlinear Phenomena},
	number = {1-3},
	pages = {77--88},
	publisher = {Elsevier},
	title = {Quantum chaos: localization vs. ergodicity},
	volume = {33},
	year = {1988}}

@article{stockmann2007quantum,
	author = {St{\"o}ckmann, Hans-J{\"u}rgen},
	journal = {Quantum Chaos},
	title = {Quantum chaos},
	year = {2007}}

@incollection{haake1991quantum,
	author = {Haake, Fritz},
	booktitle = {Quantum coherence in mesoscopic systems},
	pages = {583--595},
	publisher = {Springer},
	title = {Quantum signatures of chaos},
	year = {1991}}

@book{gutzwiller2013chaos,
	author = {Gutzwiller, Martin C},
	publisher = {Springer Science \& Business Media},
	title = {Chaos in classical and quantum mechanics},
	volume = {1},
	year = {2013}}

@article{altmann2006stickiness,
	author = {Altmann, Eduardo G and Motter, Adilson E and Kantz, Holger},
	journal = {Physical Review E---Statistical, Nonlinear, and Soft Matter Physics},
	number = {2},
	pages = {026207},
	publisher = {APS},
	title = {Stickiness in Hamiltonian systems: From sharply divided to hierarchical phase space},
	volume = {73},
	year = {2006}}

@book{devaney2018introduction,
	author = {Devaney, Robert},
	publisher = {CRC press},
	title = {An introduction to chaotic dynamical systems},
	year = {2018}}

@article{dana1989resonances,
	author = {Dana, I and Murray, NW and Percival, IC},
	journal = {Physical review letters},
	number = {3},
	pages = {233},
	publisher = {APS},
	title = {Resonances and diffusion in periodic Hamiltonian maps},
	volume = {62},
	year = {1989}}

@article{vaienti1992ergodic,
	author = {Vaienti, S},
	journal = {Journal of statistical physics},
	number = {1},
	pages = {251--269},
	publisher = {Springer},
	title = {Ergodic properties of the discontinuous sawtooth map},
	volume = {67},
	year = {1992}}

@article{wojtkowski1981model,
	author = {Wojtkowski, Maciej},
	journal = {Communications in Mathematical Physics},
	number = {4},
	pages = {453--464},
	publisher = {Springer},
	title = {A model problem with the coexistence of stochastic and integrable behaviour},
	volume = {80},
	year = {1981}}

@phdthesis{altmann2007intermittent,
	author = {Altmann, Eduardo Goldani},
	school = {Verlag nicht ermittelbar},
	title = {Intermittent chaos in Hamiltonian dynamical systems},
	year = {2007}}

@article{akaishi2009accumulation,
	author = {Akaishi, A and Shudo, A},
	journal = {Physical Review E---Statistical, Nonlinear, and Soft Matter Physics},
	number = {6},
	pages = {066211},
	publisher = {APS},
	title = {Accumulation of unstable periodic orbits and the stickiness in the two-dimensional piecewise linear map},
	volume = {80},
	year = {2009}}

@article{malovrh2002spectral,
	author = {Malovrh, Jure and Prosen, Tomaz},
	journal = {Journal of Physics A: Mathematical and General},
	number = {10},
	pages = {2483},
	publisher = {IOP Publishing},
	title = {Spectral statistics of a system with sharplydivided phase space},
	volume = {35},
	year = {2002}}

@article{lee1989makes,
	author = {Lee, Koo-Chul},
	journal = {Physica D: Nonlinear Phenomena},
	number = {1-2},
	pages = {186--202},
	publisher = {Elsevier},
	title = {What makes chaos border sticky?},
	volume = {35},
	year = {1989}}

@book{skokos2016chaos,
	author = {Skokos, Charalampos Haris and Gottwald, Georg A and Laskar, Jacques},
	publisher = {Springer},
	title = {Chaos detection and predictability},
	volume = {1},
	year = {2016}}

@article{keating1999quantum,
	author = {Keating, JP and Mezzadri, F and Robbins, JM},
	journal = {Nonlinearity},
	number = {3},
	pages = {579},
	publisher = {IOP Publishing},
	title = {Quantum boundary conditions for torus maps},
	volume = {12},
	year = {1999}}

@article{hillebrand2022quantifying,
	author = {Hillebrand, Malcolm and Zimper, Sebastian and Ngapasare, Arnold and Katsanikas, Matthaios and Wiggins, S and Skokos, Ch},
	journal = {Chaos: An Interdisciplinary Journal of Nonlinear Science},
	number = {12},
	publisher = {AIP Publishing},
	title = {Quantifying chaos using Lagrangian descriptors},
	volume = {32},
	year = {2022}}

@article{mancho2013lagrangian,
	author = {Mancho, Ana M and Wiggins, Stephen and Curbelo, Jezabel and Mendoza, Carolina},
	journal = {Communications in Nonlinear Science and Numerical Simulation},
	number = {12},
	pages = {3530--3557},
	publisher = {Elsevier},
	title = {Lagrangian descriptors: A method for revealing phase space structures of general time dependent dynamical systems},
	volume = {18},
	year = {2013}}

@article{lopesino2015lagrangian,
	author = {Lopesino, Carlos and Balibrea, Francisco and Wiggins, Stephen and Mancho, Ana M},
	journal = {Communications in Nonlinear Science and Numerical Simulation},
	number = {1-3},
	pages = {40--51},
	publisher = {Elsevier},
	title = {Lagrangian descriptors for two dimensional, area preserving, autonomous and nonautonomous maps},
	volume = {27},
	year = {2015}}

@article{saraceno1994towards,
	author = {Saraceno, Marcos and Voros, Andr{\'e}},
	journal = {Physica D: Nonlinear Phenomena},
	number = {2-4},
	pages = {206--268},
	publisher = {Elsevier},
	title = {Towards a semiclassical theory of the quantum baker's map},
	volume = {79},
	year = {1994}}

@article{leboeuf1990phase,
	author = {Leboeuf, P and Kurchan, J and Feingold, M and Arovas, DP},
	journal = {Physical review letters},
	number = {25},
	pages = {3076},
	publisher = {APS},
	title = {Phase-space localization: topological aspects of quantum chaos},
	volume = {65},
	year = {1990}}

@incollection{backer2003numerical,
	author = {B{\"a}cker, Arnd},
	booktitle = {The mathematical aspects of quantum maps},
	pages = {91--144},
	publisher = {Springer},
	title = {Numerical aspects of eigenvalue and eigenfunction computations for chaotic quantum systems},
	year = {2003}}

@article{keating1991cat,
	author = {Keating, Jonathan P},
	journal = {Nonlinearity},
	number = {2},
	pages = {309},
	publisher = {IOP Publishing},
	title = {The cat maps: quantum mechanics and classical motion},
	volume = {4},
	year = {1991}}

@article{hannay1980quantization,
	author = {Hannay, John H and Berry, Michael V},
	journal = {Physica D: Nonlinear Phenomena},
	number = {3},
	pages = {267--290},
	publisher = {Elsevier},
	title = {Quantization of linear maps on a torus-Fresnel diffraction by a periodic grating},
	volume = {1},
	year = {1980}}

@article{backer2008regular,
	author = {B{\"a}cker, A and Ketzmerick, R and L{\"o}ck, S and Schilling, L},
	journal = {Physical review letters},
	number = {10},
	pages = {104101},
	publisher = {APS},
	title = {Regular-to-chaotic tunneling rates using a fictitious integrable system},
	volume = {100},
	year = {2008}}

@article{ketzmerick2000new,
	author = {Ketzmerick, Roland and Hufnagel, Lars and Steinbach, Frank and Weiss, Matthias},
	journal = {Physical Review Letters},
	number = {6},
	pages = {1214},
	publisher = {APS},
	title = {New class of eigenstates in generic Hamiltonian systems},
	volume = {85},
	year = {2000}}

@article{podolskiy2003semiclassical,
	author = {Podolskiy, Viktor A and Narimanov, Evgenii E},
	journal = {Physical review letters},
	number = {26},
	pages = {263601},
	publisher = {APS},
	title = {Semiclassical description of chaos-assisted tunneling},
	volume = {91},
	year = {2003}}

@article{tomsovic1994chaos,
	author = {Tomsovic, Steven and Ullmo, Denis},
	journal = {Physical Review E},
	number = {1},
	pages = {145},
	publisher = {APS},
	title = {Chaos-assisted tunneling},
	volume = {50},
	year = {1994}}

@article{bohigas1993manifestations,
	author = {Bohigas, Oriol and Tomsovic, Steven and Ullmo, Denis},
	journal = {Physics Reports},
	number = {2},
	pages = {43--133},
	publisher = {Elsevier},
	title = {Manifestations of classical phase space structures in quantum mechanics},
	volume = {223},
	year = {1993}}

@article{ishii2012weyl,
	author = {Ishii, Akihiro and Akaishi, Akira and Shudo, Akira and Schomerus, Henning},
	journal = {Physical Review E---Statistical, Nonlinear, and Soft Matter Physics},
	number = {4},
	pages = {046203},
	publisher = {APS},
	title = {Weyl law for open systems with sharply divided mixed phase space},
	volume = {85},
	year = {2012}}

@article{percival1973regular,
	author = {Percival, IC},
	journal = {Journal of Physics B: Atomic and Molecular Physics},
	number = {9},
	pages = {L229},
	publisher = {IOP Publishing},
	title = {Regular and irregular spectra},
	volume = {6},
	year = {1973}}

@article{berry1977regular,
	author = {Berry, Michael V},
	journal = {Journal of Physics A: Mathematical and General},
	number = {12},
	pages = {2083},
	publisher = {IOP Publishing},
	title = {Regular and irregular semiclassical wavefunctions},
	volume = {10},
	year = {1977}}

@article{berry1984semiclassical,
	author = {Berry, Michael V and Robnik, Marko},
	journal = {Journal of Physics A: Mathematical and General},
	number = {12},
	pages = {2413},
	publisher = {IOP Publishing},
	title = {Semiclassical level spacings when regular and chaotic orbits coexist},
	volume = {17},
	year = {1984}}

@article{prosen1994semiclassical,
	author = {Prosen, Toma{\v{z}} and Robnik, Marko},
	journal = {Journal of Physics A: Mathematical and General},
	number = {24},
	pages = {8059},
	publisher = {IOP Publishing},
	title = {Semiclassical energy level statistics in the transition region between integrability and chaos: transition from Brody-like to Berry-Robnik behaviour},
	volume = {27},
	year = {1994}}
\end{document}